\documentclass[12pt]{article}

\usepackage[utf8]{inputenc}
\usepackage{bm}
\usepackage{graphicx}
\usepackage[colorinlistoftodos]{todonotes}
\usepackage[colorlinks=true, allcolors=blue]{hyperref}
\usepackage{algorithm}  
\usepackage{algpseudocode}  
\usepackage{amsmath} 
\usepackage{enumitem}
\usepackage{amsthm}
\usepackage{authblk}
\usepackage{amsfonts}
\usepackage{mathrsfs}
\usepackage{titling}
\usepackage{multirow}
\usepackage{subcaption}
\usepackage{soul}
\usepackage{verbatim}
\usepackage{amssymb}
\usepackage{bbm}
\usepackage{bm}
\usepackage{diagbox}
\newtheorem{theorem}{Theorem}

\newtheorem{assumption}{Assumption}
\newtheorem{definition}{Definition}
\theoremstyle{remark}

\usepackage{natbib}
\usepackage{multibib}
\usepackage[super]{nth}

\newcites{New}{References}

\usepackage{geometry}
\title{
\Large
A Graphical Point Process Framework for Understanding Removal Effects in Multi-Touch Attribution
}
\author{Jun Tao$^{1,2}$}
\author{Qian Chen$^1$}
\author{James W. Snyder Jr.$^2$}
\author{Arava Sai Kumar$^2$}
\author{\\Amirhossein Meisami$^2$}
\author{Lingzhou Xue$^1$}
\affil{$^1$The Pennsylvania State University; $^2$Adobe Inc.}
\date{First Version: May 2022; This Version: February 2023}

\begin{document}
\maketitle

\begin{abstract}
Marketers employ various online advertising channels to reach customers, and they are particularly interested in attribution -- measuring the degree to which individual touchpoints contribute to an eventual conversion. The availability of individual customer-level path-to-purchase data and the increasing number of online marketing channels and types of touchpoints bring new challenges to this fundamental problem. We aim to tackle the attribution problem with finer granularity by conducting attribution at the path level. To this end, we develop a novel graphical point process framework to study the direct conversion effects and the full relational structure among numerous types of touchpoints simultaneously. Utilizing the temporal point process of conversion and the graphical structure, we further propose graphical attribution methods to allocate proper path-level conversion credit, called the attribution score, to individual touchpoints or corresponding channels for each customer’s path to purchase. Our proposed attribution methods consider the attribution score as the removal effect, and we use the rigorous probabilistic definition to derive two types of removal effects. We examine the performance of our proposed methods in extensive simulation studies and compare their performance with commonly used attribution models. We also demonstrate the performance of the proposed methods in a real-world attribution application. 

\vspace{32pt}
\noindent%
\end{abstract}

\section{Introduction}

\begin{figure}[H]
    \centering
    \includegraphics[width=12cm]{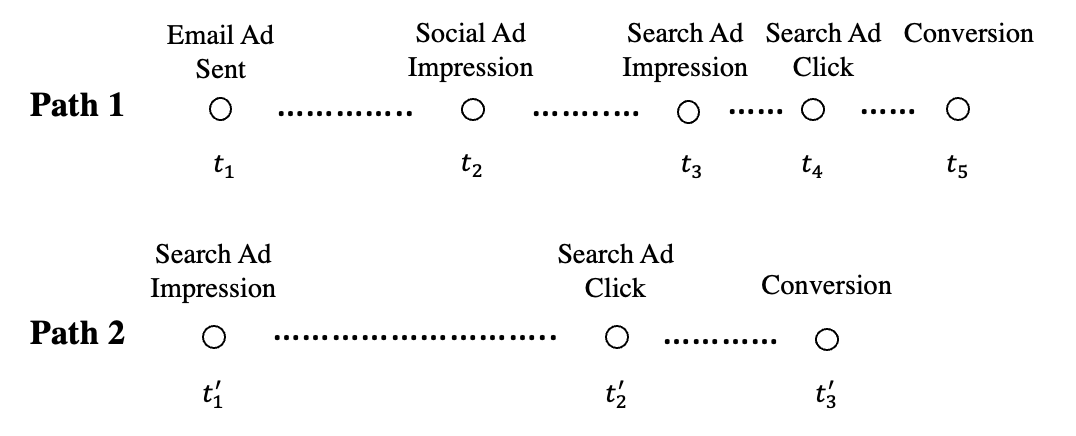}
    \caption{The illustrations of customer-level path-to-purchase data.}
    \label{fig:Path}
\end{figure}

Attribution is a classic problem. The granularity of data plays an important role in studying this problem. Early work investigating synergies across channels employed aggregate data to extract insights for customer targeting and marketing budget allocation \citep[e.g.,][]{naik2003understanding}. With advances in online data collection, individual customer-level path-to-purchase data, describing when and how individual customers interact with various channels in their purchase funnels, becomes available. Such data availability led attribution modeling to a new phase. Paths to purchases differ among customers, and the temporal distances between touchpoints also differ. In Figure~\ref{fig:Path}, we provide two illustrative examples of customer-level path-to-purchase data. As shown in Path 1, one customer received an email ad about the focal product at time $t_{1}$, and then this customer saw an ad about the product on social media at time $t_{2}$. Later, this customer searched for this product in a search engine, and its paid ad appeared at the top of the search result page at time $t_{3}$. This customer clicked on this product ad at time $t_{4}$ to visit the product's website and then purchased the product at time $t_{5}$. As shown in Path 2, another customer saw a search engine ad about this product at time $t_{1}'$ and then clicked on this ad at time $t_{2}'$ to visit the product's website and converted at time $t_{3}'$.

Researchers have started to utilize customer-level path-to-purchase data for attribution modeling \citep[e.g.,][]{xu2014path,li2014attributing, anderl2016mapping}. In recent years, we have witnessed an increasing number of online marketing channels and types of touchpoints. The channels include various search engines (e.g., Google, Bing), social media platforms (e.g., LinkedIn, Twitter, Instagram, Tiktok, Snapchat, WhatsApp), display ads, emails, web banners, app banners, customer support, desktop notification, and many others. Meanwhile, there are a lot of different types of touches within each search engine/platform/media. For example, ad impressions and ad clicks are different touchpoints that may have different conversion effects. To enable granular marketing, marketers are interested in evaluating the marketing performance of each search engine/platform/media and even the performance of a particular type of touchpoints. But in most existing research about attribution modeling, touchpoints on the path are typically aggregated to the channel they belong to (e.g., search, display, email); different search engines/platforms/media within the same channels are not differentiated, and different types of touchpoints within the same search engine/platform/media are not differentiated either. For example, both search ad impression and click are considered customers' interactions with the search channel and are not studied separately; touchpoints via Instagram, Twitter, and other social media platforms are not differentiated but studied as one social media channel as a whole. In addition, in most existing studies, only a few channels are studied simultaneously. 

Similar to recent work in attribution modeling, we also focus on the path-to-purchase data. But we aim to achieve finer granularity by conducting attribution at the path level and providing touchpoint-wise scores. Namely, we aim to allocate appropriate credit for the conversion to each touchpoint or a subset of touchpoints for each customer's path to purchase. By doing this, we can provide attribution scores for both individual touchpoints and channels, which can help marketers make better marketing decisions. Also, we would like to study the touchpoints from more channels simultaneously, which better suits firms' current needs. 

Attributing proper credit to each touchpoint is very challenging, as touchpoints interplay with each other both within-channel and across-channel. For instance, seeing email and social ads may increase a customer's future probability of searching for the product in a search engine and then clicking the search ad to visit the firm’s website and make an online purchase. Also, a touchpoint (e.g., search ad impression) may trigger the occurrence of a future touchpoint within the same channel (e.g., search ad click), leading to a conversion. The phenomenon that the earlier touches may increase the probability of the occurrence of future touches and the possible conversions are well discussed in \cite{li2014attributing} as carryover effect (through the same channels) and spillover effects (through other channels). 

To properly capture both direct and indirect conversion effects of various touchpoints, we must consider the following characteristics in attribution modeling. Firstly, different types of touchpoints may significantly vary in their probability of occurring, exciting other types of touchpoints and affecting purchase conversion. Modeling the multivariate nature of different types of touchpoints is necessary. Secondly, all marketing effects, including conversion effects and the interactive effects among touches, decay over time. It is essential to consider the time interval between two touches in attribution modeling. Solely utilizing the sequence information of the touches ignores such decaying effects, resulting in estimation biases. Thirdly, touchpoints and purchase conversions will likely gather together as clusters on the timeline. In other words, the path-to-purchase data are clumpy \citep{zhang2013new}. Fourthly, the attribution modeling and its estimation methods need to be scalable for analyzing a large number of marketing channels and types of touchpoints.

\subsection{Related Literature}

The attribution problem has been studied broadly by researchers from both academia and industry for over two decades. In this subsection, we provide a brief overview of existing approaches, and more details can be found in the survey papers \citep{ kannan2016path,gaur2020attribution}. Existing attribution models can be classified into two major categories: rule-based heuristics and data-driven approaches.  

\subsubsection{Rule-based Heuristics}
Simple rule-based heuristics are widely used for multi-touch attribution in practice. For example, the last-touch attribution method assigns the credit solely to the touchpoint directly preceding the conversion; the first-touch attribution assigns the credit to the first touchpoint in the customer journey; the U-shaped method assigns an equal amount of credit to the first and the last touchpoints while evenly distributing the remaining credit amongst the other touchpoints. These rule-based heuristics can still be easily employed for the path-to-purchase data, but they either ignore the effects of other touchpoints or do not consider the interactive effects among touchpoints. They are also criticized for being biased and lacking rationale justifying their appropriateness as attribution measures \citep{singal2022shapley}.

\subsubsection{Data-driven Approaches}
The data-driven approaches mainly consist of incremental value (or removal effect) approaches and Shapley value approaches. The main novelty of previous work in this category comes from the following two perspectives. The majority of research focuses on proposing new models to describe user behavior \citep[e.g.,][]{shao2011data,breuer2011incorporating,danaher2018delusion, xu2014path, zhao2019revenue}, while the others focus on studying the attribution scoring methods such as the justification and fairness of Sharpley value \citep[e.g.,][]{dalessandro2012causally, singal2022shapley}. 

\begin{itemize}
    \item The incremental value (or removal effect) approaches compute the change in the conversion probability when one touchpoint or a set of touchpoints are removed from a customer's path. As a result, the change in the conversion probability is also known as the removal effect. In the past decade, researchers have developed a variety of models to describe consumer behavior, such as regression models \citep[e.g.,][]{shao2011data, breuer2011incorporating, danaher2018delusion, zhao2019revenue}, Markov models \citep{yang2010analyzing, anderl2016mapping, berman2018beyond, kakalejvcik2018multichannel}, Bayesian models \citep[e.g.,][]{li2014attributing}, time series models \citep{kireyev2016display, de2016effectiveness}, survival theory-based models \citep{zhang2014multi, ji2016probabilistic}, deep learning models \citep{li2018deep, kumar2020camta}, and so on. The main novelty of previous work in this line comes from modeling user behavior. Most of them consider the touchpoints as deterministic rather than stochastic events and ignore the dynamic interactions among these marketing communications and interventions \citep{xu2014path}. However, it is important to account for the exciting effects of these touchpoints. Also, as pointed out by \cite{singal2022shapley}, there exists little (if any) theoretical justification for the attribution based on the incremental value. 
    
    \item The Shapley value approaches apply the game theory-based concept of Shapley value \citep{shapley1997value} for allocating credit to individual players in a cooperative game. Due to the nature of the Shapley value, it typically provides channel-level but not path-level attribution or touchpoint-wise attribution scores. In addition, existing methods based on Shapley value did not take into account the temporal distance between touchpoints in the path-to-purchase data, including \citep{dalessandro2012causally, de2016effectiveness, kireyev2016display, berman2018beyond, singal2022shapley}. For example, the most recent work by \cite{singal2022shapley} used a discrete Markov chain model to describe the transitions in a customer's state along the customer journey through the conversion funnel, which does not incorporate the temporal distance when the customer moves from a state to another state in one transition.

    \item Attribution has also been investigated from other angles. For example, \cite{xu2014path} proposed a Bayesian method using a multivariate point process and calculated the attribution scores using simulations. However, this simulation-based attribution method is computationally expensive and unable to provide a path-level score for each observed path with a conversion.

\end{itemize}

\subsection{Our Approach}

To the best of our knowledge, none of the existing models can study the full relational structure of numerous types of touches across multi-channels, which is essential to understanding both the direct and indirect conversion effects of each type of touch and the corresponding channels. Given the potentially large number of various types of touchpoints under study, a model that can simultaneously study the interactive and conversion effects of these many types of touchpoints is in need. To fill these research gaps, we make the following efforts in this work:

Firstly, we develop a novel graphical point process model for attribution to describe customer behavior in the multi-channel setting using customer-level path-to-purchase data. The graphical model not only learns the direct conversion effects of numerous types of touchpoints but also estimates exciting effects among different types of touches simultaneously. 

Secondly, we propose graphical attribution methods to allocate proper conversion credit, called the attribution score, to individual touchpoints and the corresponding channels for each customer’s path to purchase. Our proposed methods consider the attribution score as the removal effect. We derive two versions of the removal effect using the temporal point process of conversion and the graphical structure.

Thirdly, we propose a regularization method for simultaneous edge selection and parameter estimation. We design a customized alternating direction method of multipliers (ADMM) to solve this optimization problem in an efficient and scalable way. In addition, we provide a theoretical guarantee by establishing the asymptotic theory for parameter estimates.

In what follows, we briefly introduce the idea of our proposed graphical point process model. As the first step, we model path-to-purchase data as multivariate temporal point processes. More specifically, the proposed method considers individual paths of touchpoints as independent event streams. Each stream consists of various types of touchpoints (events) occurring irregularly and asynchronously on a common timeline. To capture the dynamic inter-dependencies (e.g., exciting patterns) among touches from a large number of independent event streams, the proposed method models event streams as multivariate temporal point processes. The multivariate temporal point processes are commonly characterized using conditional intensity functions \citep{gunawardana2011model}, which describe the instantaneous rates of occurrences of future touchpoints given the history of prior touchpoints. The proposed model can consider the temporal distances and the clumpy nature of touches. \cite{xu2014path} is the first paper to tackle attribution modeling using multivariate temporal point processes. They consider advertisement clicks and purchases as dependent random events in continuous time and cast the model in the Bayesian hierarchical framework. 

The proposed model further introduces a Granger causality graph that is a directed graph to represent the dependencies among various event types. The nodes in the graph represent event types, and directed edges depict the historical influence of one type of event on the others, which are called the Granger causality relations \citep{granger1969investigating}. The proposed graphical model allows for a large number of online marketing channels and types of touchpoints. By fully capturing the Granger causality relations among various event types, the proposed model simultaneously measures how the numerous types of touches affect conversion, as well as the exciting effects of different types of touches within and across channels.

Based on the learned graph, we propose graphical attribution methods to assign proper credit for conversions to each type of touchpoint or a corresponding channel. The conversion credit, also called attribution scores, are calculated at the path level from a granular point of view. They can be aggregated to the channel level when necessary. The first attribution method measures the direct effect of the event(s) of interest on the conversion. This is the relative change in conversion intensity when we remove only the event(s) of interest from the path and assume other touchpoints on the path remain unaffected. The second attribution method fully uses the graphical causality structure and measures the total removal effect of the event(s) of interest. The corresponding attribution score is the marginal lift of the expected intensity of conversion by considering not only the removal of the events of interest but also the potential loss of subsequent customer-initiated events.

We examine the performance of our proposed methods in simulation studies and compare their performance with commonly used attribution models using two sets of simulated data. One data set is simulated from the multivariate Hawkes process \citep{hawkes1971spectra}. The other data set is simulated from a modified version of the Digital Advertising System Simulation (DASS) developed by Google Inc. The simulated data includes online customer browsing behavior and injected advertising events that impact customer behavior. Our proposed methods outperform the benchmark models in measuring channels' contribution to conversions. Moreover, we demonstrate the performance of the proposed methods in a real-world attribution application. 

\subsection{Our Contributions}
We provide practitioners with a new attribution modeling tool to understand how different marketing efforts contribute to conversions at a finer granularity in online multi-channel settings, where there exist a potentially large number of different types of touchpoints nowadays. Our tool distributes proper credit to individual touchpoints and the corresponding channels by conducting attribution modeling at the individual customer path level. Our tool helps firms' granular marketing operations, budget allocation, profit maximization, etc. 

In addition to the substantive contribution, we have the following methodological contributions to the literature.

Firstly, we contribute to the attribution modeling literature by proposing a graphical point process model to describe customer behavior using individual customer-level path-to-purchase data. We rigorously model the exciting effects of numerous types of touches in this framework and develop an efficient penalized algorithm for model estimation. We apply Granger causality in a marketing context to study the temporal relations among marketing activities. We also establish the asymptotic theory for parameter estimates.
       
Secondly, our proposed graphical attribution methods contribute to the literature on the incremental value (or removal effect) approaches. We provide a rigorous probabilistic definition of attribution scores and derive two types of removal effects, namely, the direct and total removal effects. We develop a new efficient thinning-based simulation method and a backpropagation algorithm for the calculation of two types of removal effects, respectively.

We organize the rest of the paper as follows. We introduce the proposed graphical point process model in Section 2 and the proposed graphical attribution methods in Section 3. We present the model estimation, computational details, and asymptotic properties in Section 4. We then demonstrate the performance of the proposed method and algorithm with the simulated data in Section 5 and provide an empirical application in Section 6. We conclude this work in Section 7.

\section{Graphical Point Process Model}
To tackle the attribution problem, we first propose a graphical point process model. This model utilizes customer-level path-to-purchase data to learn the full relational structure among different types of touchpoints. This section introduces the proposed graphical point process model.

Our graphical point process model considers the observed individual paths to purchases as independent event streams, where each stream consists of various types of events (i.e., touchpoints and conversions) occurring irregularly and asynchronously on a common timeline. Suppose there are $p$ unique types of events, which can be labeled as $1,\dots,p$. A customer's path $D$ is represented by $\{(t_i,e_i)\}_{i=1}^m$ with $0\le t_1<t_2<\dots<t_m\le T$, where $m$ is the total count of occurred events and $T$ is the length of observation. For the $i$-th event $(t_i,e_i)$, $t_i$ is its timestamp, and $e_i\in\mathcal{E}=\{1,2,\dots,p\}$ is its event label which specifies the type of touchpoint (e.g., social ad impression, social ad click, email sent, email ad click, search ad impression, search ad click, display ad impression, display ad click) or conversion. Without loss of generality, let $e=1$ denote the label of conversion. Given the path $D$, if there exists $i\le m$ such that $e_i=1$, which means there is a conversion event, then such a path $D$ is called a positive path. Otherwise, $D$ is called a negative path. Suppose we observe $n$ paths, $D_{1},\dots,D_{n}$, where $D_{j}=\{(t_i^j,e_i^j)\}_{i=1}^{m_j}$ is the $j$-th path with length $T_{j}$. 

To capture the dynamic inter-dependencies among touches from a large number of independent event streams, the proposed framework models event streams as multivariate temporal point processes. It introduces a directed Granger causality graph to represent the dependencies among various types of touchpoints in the event streams. In this section, we first provide a brief overview of the temporal point process and the Granger causality graph. Then we will introduce our proposed model in detail.

\subsection{Temporal Point Process}
For an event type labeled as $e$, we can describe its occurrence on the timeline as a temporal point process $N_e(t)$. The function $N_e(t)$ is the number of type-$e$ events that happened until time $t$, which is a right-continuous and non-decreasing piece-wise function $\mathbb{R}_{\ge 0}\rightarrow \mathbb{N}$:
\[N_e(t) = \sum_{i:e_i=e}\mathbbm{1}_{\{t_i\le t\}}.\]
We assume $N_e(t)$ has almost surely step size $1$ and does not jump simultaneously.
If $t' < t$, then $N_e(t)-N_e(t')$ is the number of type-$e$ events that occurred during the interval $(t', t ]$, which can also be denoted by $N_e((t', t ])$. Putting $p$ event types together, we let $\mathbf{N}(t)$ denote the vector of counting functions $(N_1(t),\dots,N_p(t))^\top$. The coordinate $N_e(t)$, $e=1,\dots,p$, is characterized by its conditional intensity
\[\lambda_e (t\mid \mathcal{H}_t):=\lim _{\Delta t\to 0^+}{\frac {1}{\Delta t}}\mathbbm{P}(N_e(t+\Delta t)-N_e(t)>0 \mid \mathcal{H}_t),\]
which describes the instantaneous rates of occurrence of future type-$e$ events. The filtration $\mathcal{H}_t:= \sigma\{\mathbf{N}(u):u<t\}$ is the $\sigma$-algebra of all events up to but excluding $t$, referring to the historical information before time $t$.

An example is the multivariate Hawkes process \citep{hawkes1971spectra} with the conditional intensity function
\[\lambda_e(t\mid \mathcal{H}_t) = \mu_e+\sum_{e'=1}^p \int_0^t h_{e'e}(t-u) dN_{e'}(u).\]
Here $\mu_e>0$ is the baseline intensity serving as a background rate of occurrence regardless of historical impact, and $h_{e'e}(\cdot)\ge 0$ is called an impact function. Intuitively, each of the past $e'$-events with $h_{e'e}(\cdot)>0$ has a positive contribution to the occurrence of the current type-$e$ event by increasing the conditional intensity of event type $e$, and this influence may decrease through time. Such positive influence is called an exciting effect from event type $e'$ to event type $e$.
\subsection{Granger Causality Graph}
Let $\mathcal{E}=\{1,2,\dots,p\}$ be the set of various types of labeled events (i.e., touchpoints and conversions), whose historical influences on each other are of great interest. We use the Granger causality relations \citep{granger1969investigating} to describe the temporal dependencies among the studied types of touchpoints. If the history of event type $e'$ helps to predict event type $e$ above and beyond the history of event type $e$ alone, event type $e'$ is said to ``Granger-cause" event type $e$. The Granger causality was introduced and discussed in the original paper \cite{granger1969investigating} and follow-up papers \citep{granger1980testing,granger1988some}, while \cite{sims1972money} gave an alternative definition of Granger causality. 

 For any event label subset $W\subseteq \mathcal{E}$, let $\mathbf{N}_{W}(t)$ be the subprocess $(N_{e}(t))_{e\in W}$. For example, $\mathbf{N}_{\mathcal{E}\setminus\{e,e'\}}(t)$ means the subprocess of all the event types other than $e$ and $e'$. For a temporal point process, the Granger causality is defined below.

\begin{definition}(Local independence \citep{didelez2008graphical})
The temporal point process $N_e(t)$ is locally independent on $N_{e'}(t)$ given $\mathbf{N}_{\mathcal{E}\setminus\{e,e'\}}(t)$, denoted by $N_{e'}\nrightarrow N_e\mid \mathbf{N}_{\mathcal{E}\setminus\{e,e'\}}$, if the conditional intensity $\lambda_e(t)$ is measurable with respect to $\sigma\{\mathbf{N}_{\mathcal{E}\setminus\{e,e'\}}(u):u<t\}$ for all $t<T$. Otherwise, $N_e(t)$ is said to be locally dependent on $N_{e'}(t)$ given $\mathbf{N}_{\mathcal{E}\setminus\{e,e'\}}(t)$ with respect to $\mathcal{H}_t$, or $N_{e'}\rightarrow N_e\mid \mathbf{N}_{\mathcal{E}\setminus\{e,e'\}}$.
\end{definition}
The above definition was introduced by \cite{didelez2008graphical} for marked point processes in an interval $[0, T]$. \cite{eichler2012graphical} studied the stationary multivariate point processes like multivariate Hawkes processes in $\mathbb{R}$ and used the notion ``Granger non-causality". The above definition is equivalent to saying the temporal point process $N_{e'}(t)$ does not Granger-cause $N_{e}(t)$ with respect to historical information $\mathcal{H}_t$. Otherwise, $N_{e'}(t)$ Granger-causes $N_{e}(t)$ with respect to $\mathcal{H}_t$. For ease of interpretation, we also say that event type $e'$ Granger-causes event type $e$ in an unambiguous manner.

\begin{figure}[H]
\centering
\includegraphics[width=6cm]{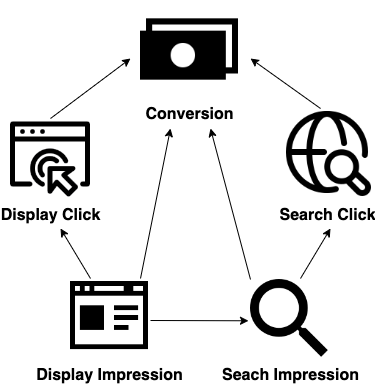}
\caption{A toy example of the Granger causality graph for attribution. All four types of touchpoints Granger-cause conversion. Within either the display channel or the search channel, impression Granger-causes click. Between the two channels, display impression Granger-causes search impression.}\label{simgraph_hawkes}
\end{figure}

We use a directed graph $\mathcal{G} = (\mathcal{E},\mathcal{A})$ like Figure~\ref{simgraph_hawkes} to represent the temporal dependencies among various types of touchpoints in the conversion paths. The node set of the graph represents the set of event types $\mathcal{E}$. The edge set $\mathcal{A}\subseteq \mathcal{E}\times \mathcal{E}$ represents the Granger causality relations between the various event types. If $N_{e'}(t)$ Granger-causes $N_{e}(t)$ with respect to $\mathcal{H}_t$, then the directed edge from node $e'$ to node $e$, denoted by $(e'\rightarrow e)$, is said to belong to $\mathcal{A}$. Such a graph $\mathcal{G} = (\mathcal{E},\mathcal{A})$ is called a Granger causality graph. For a multivariate temporal point process, the Granger causality graph is defined below.

\begin{definition}(Granger causality graph \citep{didelez2008graphical,eichler2012graphical})
A multivariate temporal point process $\mathbf{N}(t)$ is said to follow the Granger causality graph $\mathcal{G} = (\mathcal{E},\mathcal{A})$ if for any pair of nodes $e,e'\in \mathcal{E}$,
\[(e'\rightarrow e)\notin \mathcal{A} \Longleftrightarrow N_{e'}\nrightarrow N_e\mid \mathbf{N}_{\mathcal{E}\setminus\{e,e'\}}.\]
\end{definition}

\subsection{The Proposed Graphical Point Process Model}

We treat each path to purchase as a $p$-dimensional point process $\mathbf{N}(t)$. In the attribution problem, there are two categories of event types, firm-initiated event types and customer-initiated event types. The firm-initiated event types \citep{wiesel2011practice} are initiated by firms such as email sent, social ad impression, and display impression. The customer-initiated event types \citep{bowman2001managing,wiesel2011practice} are initiated by customers or prospective customers, including conversion, search ad impression and click, email click, and so forth. Let $\mathcal{E}_\mathrm{f}$ and $\mathcal{E}_\mathrm{c}$ denote the set of firm-initiated event types and the set of customer-initiated event types and thus $\mathcal{E}=\mathcal{E}_\mathrm{f}\cup\mathcal{E}_\mathrm{c}$ and $\mathcal{E}_\mathrm{f}\cap\mathcal{E}_\mathrm{c}=\emptyset$. Suppose there are $q$ customer-initiated event types for some $1\le q< p$. Without loss of generality, let $\mathcal{E}_\mathrm{c}=\{1,\dots,q\}$ with conversion labeled as $e=1$ and $\mathcal{E}_\mathrm{f}=\{q+1,\dots,p\}$. We assume that the point process of the firm-initiated event types $\mathbf{N}_{\mathcal{E}_\mathrm{f}}(t)$ is controlled by the firms or ad advertisers strategically. In other words, $\mathbf{N}_{\mathcal{E}_\mathrm{f}}(t)$ only serves as an observed input whose conditional intensity function does not require learning. For the point process of customer-initiated event types $\mathbf{N}_{\mathcal{E}_\mathrm{c}}(t)$, we model it through the following conditional intensity:
\begin{equation}\label{intensity}
    \lambda_e(t\mid \mathcal{H}_t) = \mu_e+\sum_{e'=1}^p\alpha_{e'e} \int_0^t \psi_{e'e}(t-u) dN_{e'}(u), \quad\text{for}\ e\in\mathcal{E}_\mathrm{c},
\end{equation}

\begin{itemize}
\item $\mu_e\ge 0$ is the baseline intensity, which corresponds to the sources of the occurrence of type-$e$ events other than the occurrence history of itself and other types of touchpoints, namely the intrinsic tendency of the occurrence. 
\item $\psi_{e'e}(\cdot)\ge 0,\ e'=1,\dots,p$, is a bounded, left-continuous kernel function, $\psi_{e'e}(t)=0$ when $t\le 0$ and
$\int_0^\infty\psi_{e'e}(t)dt= 1.$ The kernel function $\psi_{e'e}(\cdot)$ describes the shapes of the touchpoints' impact.
For example, $\psi_{e'e}(t)=\frac{1}{T_0}\cdot\mathbbm{1}_{\{0<t\le T_0\}}$ accounts for a constant impact of a previous type-$e'$ event within $T_0$ of its occurrence; $\psi_{e'e}(t)=\frac{1}{T_0}\exp(-\frac{t}{T_0})\cdot\mathbbm{1}_{\{t>0\}}$ works for an exponential decaying impact; $\psi_{e'e}(t)=\sqrt{\frac{2}{\pi T_0^2}}\exp(-\frac{t^2}{2T_0^2})\cdot\mathbbm{1}_{\{t>0\}}$ can be used when the decaying impact is even faster.
    \item $\alpha_{e'e}\ge 0,\ e'=1,\dots,p$, is the Granger causality coefficient. The value of $\alpha_{e'e}$ describes the scale of the temporal dependence.
\end{itemize}
To interpret the coefficient $\alpha_{e'e}$, we introduce the following theorem.

\begin{theorem} 
Assume a point process $\mathbf{N}(t)$ with conditional intensity functions defined in \eqref{intensity} follows the Granger causality graph $\mathcal{G} = (\mathcal{E},\mathcal{A})$. For any event labels $e'\in\mathcal{E}$ and $e\in\mathcal{E}_\mathrm{c}$, if the condition $N_{e'}(T)>0$ holds, then \[(e'\rightarrow e)\notin \mathcal{A}\Longleftrightarrow \alpha_{e'e}= 0.\]
\end{theorem}

This result is an adaptation from the case of the multivariate Hawkes process \citep{eichler2012graphical,xu2016learning}. Analogous to the multivariate Hawkes process, the meaning of $\alpha_{e'e}$ can be described in two cases: $\alpha_{e'e}>0$ implies an exciting effect from event type $e'$ to event type $e$ by increasing its conditional intensity $ \lambda_e(t\mid \mathcal{H}_t)$; the case $\alpha_{e'e}=0$ implies no Granger causality from event type $e'$ to event type $e$ since $\lambda_e(t\mid \mathcal{H}_t)$ is not affected by the occurrence of any type-$e'$ event. By such construction, we can use the matrix $A = \left(\alpha_{e'e}\right)_{e'\in \mathcal{E},e\in\mathcal{E}_c}\in \mathbb{R}_{\ge 0 }^{p\times q}$ to stand for the graphical Granger causality structure for customer-initiated event types.

As the firm-initiated event types are fully controlled by marketers or ad advertisers, in our proposed model, we assume that they are not dependent on other types of events. That is to say, the ground truth of $\mathcal{A}$ for $e\in \mathcal{E}_\mathrm{f}$ is known with $N_{e'}\nrightarrow N_e\mid \mathbf{N}_{\mathcal{E}\setminus\{e,e'\}}$ for any $e'\in \mathcal{E}$. 



\section{Graphical Attribution Method}
The goal of attribution is to assign proper credit for conversions to each type of touchpoint or a corresponding channel. We adopt the path-level scoring approach to obtain a granular view. This approach does not rely on the distribution of the firm-initiated event types $\mathbf{N}_{\mathcal{E}_\mathrm{f}}(t)$. So it can work for cases where customers are treated with different advertising strategies.

The path-level credit, called the attribution score, represents the potential fractional loss of a conversion on a path given the absence of certain event(s). We will refer to this value by the removal effect of the event(s). In this section, we first derive a version of the removal effect using the temporal point process of conversion -- direct removal effect by analyzing the incremental contribution of each touchpoint. Then we fully use the graphical structure and propose another version of the removal effect -- total removal effect -- which explains the marginal increase in the chance of conversion.

\subsection{Direct Removal Effect}
In this subsection, we calculate attribution scores using the direct removal effect from the point of view of the graphical point process. We define the \textit{direct removal effects} (DRE) as the relative change in conversion intensity when we remove only the event(s) of interest from the path and assume other touchpoints on the path remain unaffected. That is, the direct removal effect considers merely the influence of the studied event(s) on an occurrence of conversion and ignores their influence on other events. Graphically speaking, the direct removal effect focuses on the direct Granger causality parent nodes of conversion and does not depend on the hierarchy beyond them.


For a positive path $D$, suppose there is a conversion at $t=t_{i^\star}$ for some $1<i^\star\le m$, that is, $e_{i^\star}=1$. Let $F^{W}_{t}(D):=\{(t_i,e_i)\in D:t_i<t,\ e_i\in W\}$ be the set of occurred events before $t$ whose event labels belong to $W$, where $W\subseteq \mathcal{E}$ is an arbitrary event label subset. Especially, let $F_{t}(D)$ stand for $F^\mathcal{E}_{t}(D)$, which is the truncated path before $t$. We are interested in the influence of a subset of touchpoints $R\subseteq F_{t^\star}(D)$ on this conversion and call $R$ a removal set. Let $\mathbf{N}^D(t)=(N_1^D(t),\dots,N_p^D(t))^\top$ denote the point process with respect to path $D$ with $N_e^D(t) = \sum_{(t_i,e_i)\in D: e_i=e}\mathbbm{1}_{\{t_i\le t\}}$ for $e=1,\dots,p$. Let $\mathcal{H}_{t}^D$ be the corresponding filtration $\sigma\{\mathbf{N}^D(u):u<t\}$. We can calculate the attribution score as the direct removal effect of $R$ with respect to path $D$:
\begin{equation}\label{dre}
  \mathrm{att}_{ t_{i^\star}}^{\mathrm{(direct)}}(R\mid D):=\frac{\lambda_1( t_{i^\star}\mid \mathcal{H}_{ t_{i^\star}}^{D})-\lambda_1( t_{i^\star}\mid \mathcal{H}_{ t_{i^\star}}^{D\setminus R})}{\lambda_1( t_{i^\star}\mid \mathcal{H}_{ t_{i^\star}}^{D})}.
\end{equation}
The above definition is a general statement that does not depend on any assumptions of point process modeling. 

\begin{figure}[H]
\begin{center}
\includegraphics[width=14cm]{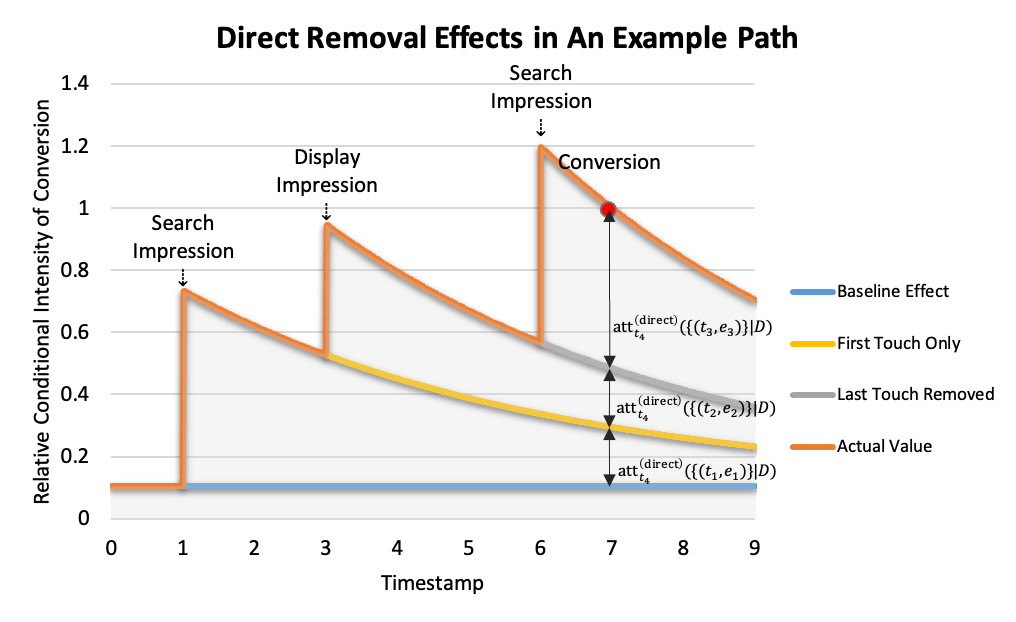}
\end{center}
\caption{The direct removal effects and the decomposition of the relative conditional intensity of conversion for an example path. The path contains three touchpoints, with $t_1=1$, $t_2=3$, $t_3=6$ and $e_1,e_2,$ and $e_3$ representing search impression, display impression, and search impression, respectively. There is a conversion at $t_4=7$. }\label{toy}
\end{figure}

Suppose that the conversion event satisfies model~\eqref{intensity}, which means the conditional intensity function of this conversion takes the form
\begin{equation}\label{conversion_intensity}
  \lambda_1( t_{i^\star}\mid \mathcal{H}_{ t_{i^\star}}) = \mu_1+\sum_{i<  i^\star}\alpha_{e_i1}\psi_{e_i1}( t_{i^\star}-t_{i}).  
\end{equation}
We can derive more specific attribution scores based on Equation \eqref{dre} and \eqref{conversion_intensity}. For example, the direct removal effect of touchpoint $(t_i,e_i)$ can be calculated by
\[\mathrm{att}_{ t_{i^\star}}^{\mathrm{(direct)}}(\{(t_i,e_i)\}\mid D) =\frac{\alpha_{e_i1} \psi_{e_i1}( t_{i^\star}-t_i)}{\lambda_1( t_{i^\star}\mid \mathcal{H}_{ t_{i^\star}}^D)}.\]
This expression shows the relationship between the Granger causality graph and the attribution score. A touchpoint $(t_i,e_i)$ can be attributed with a score only if its touchpoint type is a parent node of the conversion node on the graph or $(e_i\rightarrow 1) \in \mathcal{A}$. Generally, for a subset $R$ of $F_{t^\star}(D)$, its direct removal effect is
\begin{equation}
\mathrm{att}_{ t_{i^\star}}^{\mathrm{(direct)}}(R\mid D)=\sum_{(t_i,e_i)\in R}\frac{\alpha_{e_i1} \psi_{e_i1}( t_{i^\star}-t_i)}{\lambda_1( t_{i^\star}\mid\mathcal{H}_{ t_{i^\star}}^D)}.\label{dre_R}
\end{equation}
Besides, under model~\eqref{intensity}, the baseline effect can be defined to be $\frac{\mu_1}{\lambda_1( t_{i^\star}\mid\mathcal{H}_{ t_{i^\star}}^D)}$.
For example, Figure~\ref{toy} shows the direct removal effects of touchpoints with respect to a path consisting of three touchpoints and a conversion.

\begin{figure}[H]
\begin{center}
\includegraphics[width=12cm]{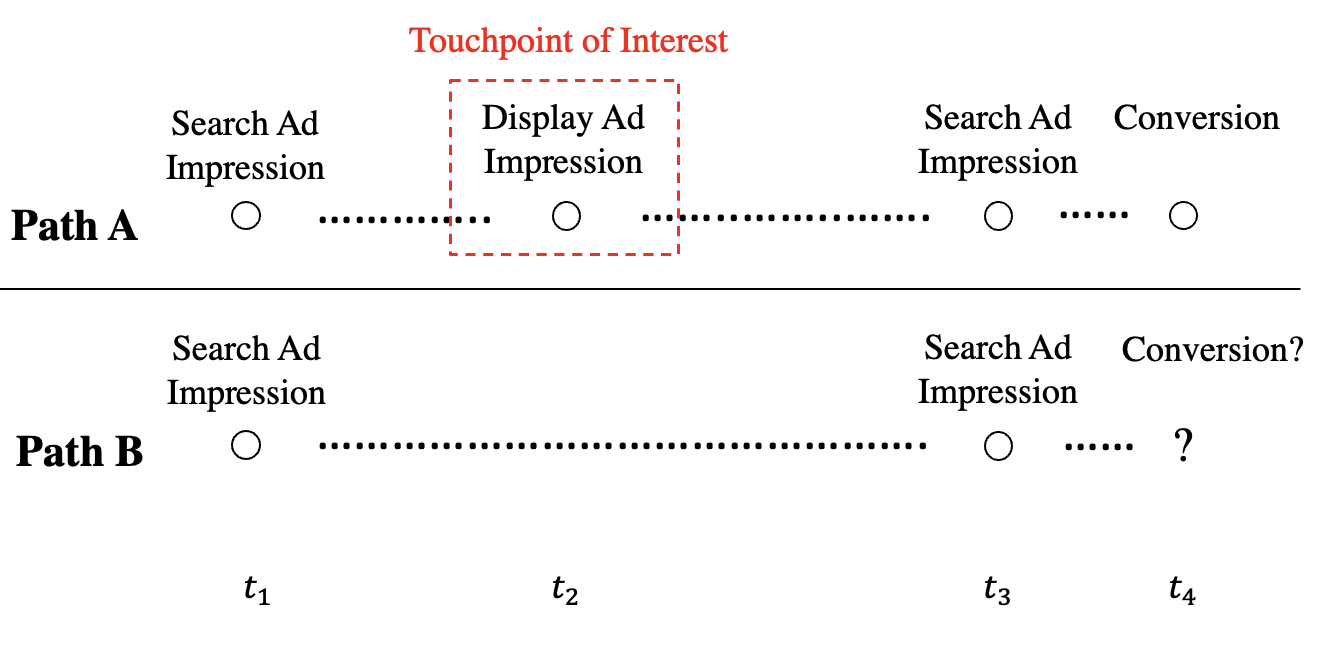}
\end{center}
\caption{The problem explained by the direct removal effect of the display impression with respect to a path containing three touchpoints. With the removal of the display impression, the direct removal effect describes the chance to see a conversion with the same timestamp.}\label{toy_RE}
\end{figure}

In the following context, we will explain the attribution score's general form \eqref{dre} in probability language. Consider an ideal experiment of two customers, A and B, with a toy example shown in Figure~\ref{toy_RE}. Suppose A and B react to touchpoints in the same way. Their conditional intensity functions for customer-initiated event types follow the same model. We observe $D_{\mathrm{A}}$, the path of A, with a conversion at $t= t_{i^\star}>0$. Let $D_{\mathrm{B}}$ denote the path of B. Suppose $F^{\mathcal{E}\setminus\{1\}}_{t_{i^\star}}(D_{\mathrm{B}})\subsetneqq F^{\mathcal{E}\setminus\{1\}}_{t_{i^\star}}(D_{\mathrm{A}})$, the touchpoints on $D_{\mathrm{B}}$ are a subset of those on $D_{\mathrm{A}}$. Assume that there is no information about $N_1^{D_{\mathrm{B}}}(t)$. Namely, we do not observe whether B converts or not. We can sample the process $N_1^{D_{\mathrm{B}}}(t)$ to obtain a complete path $D_{\mathrm{B}}$. We exploit the thinning operation for a temporal point process, which uses some definite rule to delete points of a basic point process, yielding a new point process.

\begin{theorem}\label{thinning} \citep{lewis1979simulation,ogata1981lewis} Assume a univariate temporal point process $N(t)$ in $[0,T]$ with intensity function $\lambda(t)$.
Let $t_1,\dots,t_{N(T)}$ be the timestamps of $N(t)$.
There exists a function $\lambda'(t)$ satisfying a.s.
\[\lambda'(t)\le\lambda(t),\quad\text{for }\ 0<t<T.\]
For $i=1,\dots, N(T)$, delete the point at $t=t_i$ with probability $1-\lambda'(t_i)/\lambda(t_i)$. Then the remaining points form a point process $N'(t)$ in the interval $[0,T]$ with intensity function $\lambda'(t)$.
\end{theorem}

We refer to the point process $N(t)$ as a background process for the desired point process $N'(t)$. In general, \cite{ogata1981lewis} suggested using a Poisson process as the background process for point process simulation. To sample the $N_1^{D_{\mathrm{B}}}(t)$, we can look for a constant $\overline{\lambda}$ satisfying $\overline{\lambda}\ge \lambda_1(t\mid\mathcal{H}_t^{D_\mathrm{B}})$ for $t\in [0,T]$ and use the Poisson process with rate $\overline{\lambda}$ as the background process.

With a closer look at the problem, we can find that sampling a Poisson process as the background process is not necessary. The point process of conversion for A, $N_1^{D_{\mathrm{A}}}(t)$, can also serve as a background process.
Based on \eqref{conversion_intensity}, we have $\lambda_1(t\mid\mathcal{H}_t^{D_\mathrm{B}})\le\lambda_1(t\mid\mathcal{H}_t^{D_\mathrm{A}})$ before any occurrence of conversion for B. Theorem~\ref{thinning} tells us that if we delete the conversion at $t=t_{i^\star}$ in $N_1^{D_{\mathrm{A}}}(t)$ with probability $1-\lambda_1( t_{i^\star}\mid\mathcal{H}_ {t_{i^\star}}^{D_\mathrm{B}})/\lambda_1( t_{i^\star}\mid\mathcal{H}_{t_{i^\star}}^{D_\mathrm{A}})$, then we obtain a process following the distribution described in \eqref{conversion_intensity} for B. Compared with \eqref{dre}, this probability is nothing but the direct removal effect of $D_\mathrm{A}\setminus D_\mathrm{B}$ with respect to path $D_\mathrm{A}$.

\subsection{Total Removal Effect}

The direct removal effect of a given subset of events describes the expected loss of conversion by comparing the path without it to the original path. In this subsection, we study another attribution score that measures the overall influence resulting from removing the subset of events along the path. The corresponding attribution score is the marginal lift of the expected intensity of conversion by considering not only the removal of the events of interest but also the potential loss of other related customer-initiated events.

For example, the direct removal effect of the first touchpoint $(t_1,e_1)\in D$ can be studied through $D\setminus\{(t_1,e_1)\}$. But removing such a touchpoint in the early stage of the path will result in more than the removal of itself. Some later touchpoints may get affected and tend not to occur. So $D\setminus\{(t_1,e_1)\}$ may not well represent a possible remaining path described by model~\eqref{intensity} because $\mathbf{N}_{\mathcal{E}_\mathrm{c}}^{D\setminus\{(t_1,e_1)\}}(t)$ is not guaranteed to follow the same model as $\mathbf{N}_{\mathcal{E}_\mathrm{c}}^{D}(t)$. In other words, the direct removal effect may not well reflect the overall influence of the removal of the touchpoint.

By model~\eqref{intensity}, for any $e\in\mathcal{E}_\mathrm{c}$ (including conversion),
\[\lambda_e(t\mid\mathcal{H}_{t}^{D\setminus R})\le\lambda_e(t\mid\mathcal{H}_{t}^D),\quad t\in (0,T).\]
The removal of $R$ from $D$ may result in the missing of certain subsequent touchpoints as well as paid conversion. The actual remaining path may contain even fewer event occurrences, which yields a subset of $D\setminus R$. This inspires us to perform the thinning operation to $\mathbf{N}_{\mathcal{E}_\mathrm{c}}(t)$ according to Theorem~\ref{thinning} instead of thinning the univariate process $N_1(t)$ only.

\begin{figure}[H]
\begin{center}
\includegraphics[width=12cm]{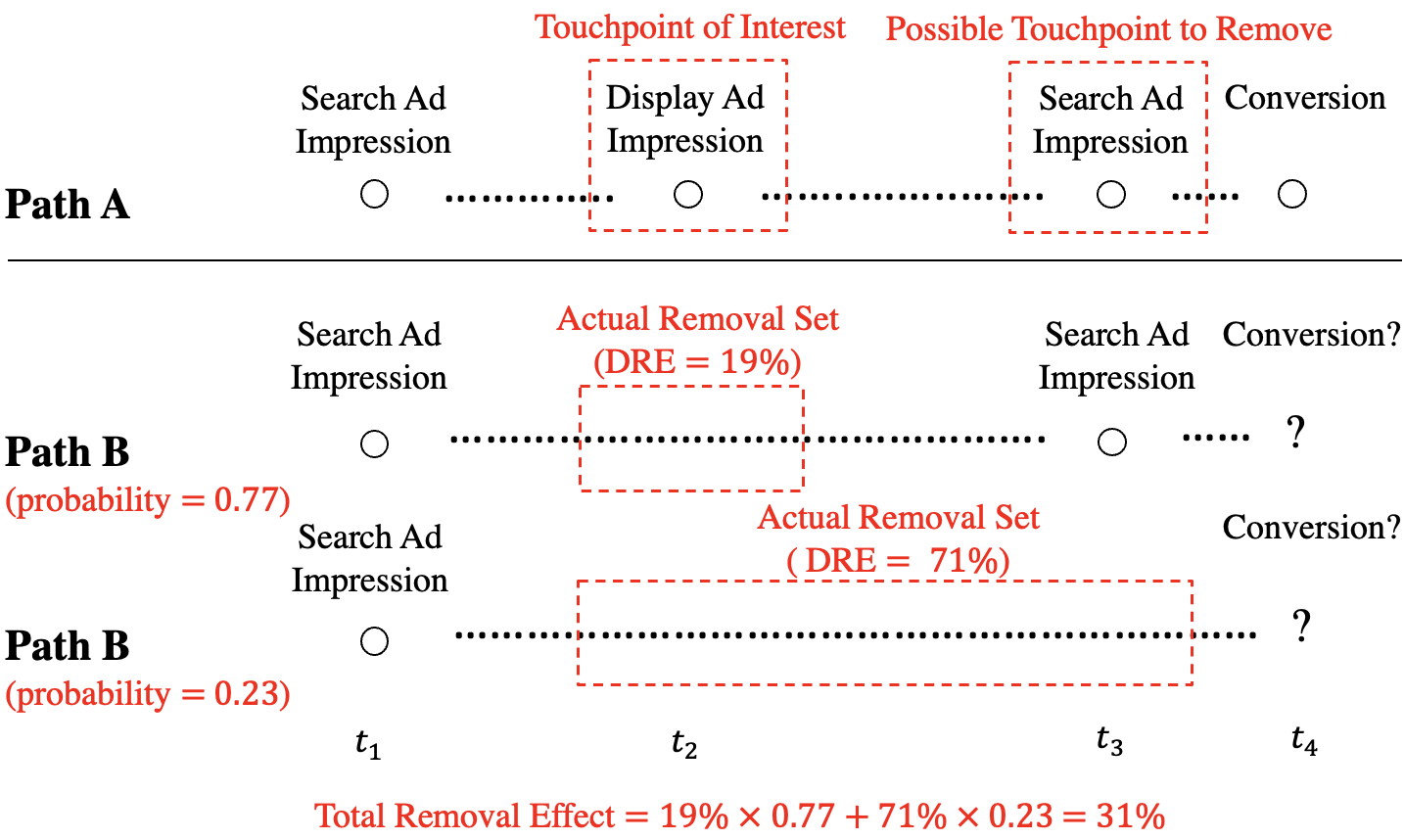}
\end{center}
\caption{For the example path containing three touchpoints, the total removal effect of the display impression can be viewed as the expected direct removal effect, where the uncertainty lies in the actual removal set. The removal of the display impression may result in the removal of the subsequent search impression.}\label{toy_TRE}
\end{figure}

 Let $i_{\min} (R) = \min\{1\le i\le m:(t_i,e_i)\in R\}$ be the event index of the first event in $R$. Recall the ideal experiment of two customers, A and B. Now we assume that $D_{\mathrm{A}}=D$ for A and the process of the customer-initiated event types $\mathbf{N}_{\mathcal{E}_\mathrm{c}}^{D_{\mathrm{B}}}(t)$ is unknown for B. Let $D_{\mathrm{B}}=D\setminus R$ first. By Theorem~\ref{thinning}, if we delete the touchpoint $(t_i,e_i)$ in $F^{\mathcal{E}_\mathrm{c}}_{t_{i^\star}}(D_{\mathrm{B}})$ from $D_{\mathrm{B}}$ with probability $1-\lambda_{e_i}(t_i\mid\mathcal{H}_{t_i}^{D_\mathrm{B}})/\lambda_{e_i}(t_i\mid\mathcal{H}_{t_i}^{D_\mathrm{A}})$ sequentially for $i>i_{\min} (R)$, then the obtained process $\mathbf{N}_{\mathcal{E}_\mathrm{c}}^{D_{\mathrm{B}}}(t)$ follows model~\eqref{intensity}. Suppose $D\setminus R^\diamond$ is the thinned version of $D_{\mathrm{B}}$ with $R^\diamond \supseteq R$ being the actual removal set. Figure~\ref{toy_TRE} shows a toy example of this idea.
As a result, given a subset $R$ of a path $D$, its \textit{total removal effect} (TRE) is defined by
\[\mathrm{att}_{ t_{i^\star}}^{\mathrm{(total)}}(R\mid D):=\mathbbm{E}[\mathrm{att}_{ t_{i^\star}}^{\mathrm{(direct)}}(R^\diamond\mid D)\mid D]= \frac{\lambda_1( t_{i^\star}\mid\mathcal{H}_{ t_{i^\star}}^D)-\mathbbm{E}[\lambda_1( t_{i^\star}\mid\mathcal{H}_{ t_{i^\star}}^{D\setminus R^\diamond})\mid D]}{\lambda_1( t_{i^\star}\mid\mathcal{H}_{ t_{i^\star}}^D)},\]
where the conditional expectation is with respect to the actual removal set $R^\diamond$ from the random thinning operation. The algorithm with the thinning operation is summarized in Algorithm~\ref{algo2}.

\begin{algorithm}[H]
  \caption{Total removal effect: thinning}
  \begin{algorithmic} 
\State Input: A path $D$, a specified conversion timestamp $ t_{i^\star}$ ($i^\star>1$) and a nonempty removal set $R\subseteq F_{t^\star}(D)$. Model parameters $\bm{\mu},\bm{\alpha}_1,\dots,\bm{\alpha}_q$ and a large integer $L$.
\For {$\ell=1,\dots,L$} 
\State $R^\diamond = R$.
      \For {$i$ in $\{i> i_{\min} (R):(t_i,e_i)\in F^{\mathcal{E}_\mathrm{c}}_{t_{i^\star}}(D\setminus R)\}$ (ascending order)} 
      \State Update $R^\diamond = R^\diamond\cup\{(t_i,e_i)\}$ with probability $1-\frac{\lambda_{e_i}(t_i\mid \mathcal{H}_{t_i}^{D\setminus R^\diamond})}{\lambda_{e_i}(t_i\mid \mathcal{H}_{t_i}^D)}$.
    \EndFor
     \State $x_{\ell} = \mathrm{att}_{ t_{i^\star}}^{\mathrm{(direct)}}(R^\diamond\mid D)$.
 \EndFor
\State    
Return: $\mathrm{att}_{ t_{i^\star}}^{\mathrm{(total)}}(R\mid D)=\frac{1}{L}\sum_{\ell=1}^L x_{\ell}$.
 \end{algorithmic} 
  \label{algo2}
\end{algorithm}

The largest value of $R^\diamond$ is $R\cup \{(t_i,e_i)\in F^{\mathcal{E}_\mathrm{c}}_{t_{i^\star}}(D\setminus R):i> i_{\min} (R)\}$, denoted by $\Omega$ for ease of notation.
Using the thinning algorithm, we can derive an explicit form of the total removal effect:
\[\mathrm{att}_{ t_{i^\star}}^{\mathrm{(total)}}(R\mid D) =\sum_{R\subseteq R' \subseteq\Omega}\mathrm{att}_{ t_{i^\star}}^{\mathrm{(direct)}}(R'\mid D)\mathbbm{P}(R^\diamond=R'\mid D). \]
During the thinning operation, the value of $R^\diamond$ depends on sampling a sequence of Bernoulli random variables. 
Let $U_i\sim \mathrm{Bernoulli}(1-\frac{\lambda_{e_i}(t_i\mid \mathcal{H}_{t_i}^{D\setminus R'})}{\lambda_{e_i}(t_i\mid \mathcal{H}_{t_i}^D)})$ be the Bernoulli random variable for thinning event $(t_i,e_i)\in\Omega$.
Then the conditional probability mass function of $R^\diamond$ is
\begin{align*}
    \ \mathbbm{P}(R^\diamond=R'\mid D)
    =& \prod_{(t_i,e_i)\in \Omega\setminus R}\mathbbm{P}(U_i=\mathbbm{1}_{\{(t_i,e_i)\in R'\}})\\
    =& \prod_{(t_i,e_i)\in \Omega\setminus R}\left(1-\frac{\lambda_{e_i}(t_i\mid \mathcal{H}_{t_i}^{D\setminus R'})}{\lambda_{e_i}(t_i\mid \mathcal{H}_{t_i}^D)}\right)^{\mathbbm{1}_{\{(t_i,e_i)\in R'\}}}\left(\frac{\lambda_{e_i}(t_i\mid \mathcal{H}_{t_i}^{D\setminus R'})}{\lambda_{e_i}(t_i\mid \mathcal{H}_{t_i}^D)}\right)^{1-\mathbbm{1}_{\{(t_i,e_i)\in  R'\}}}.
\end{align*}

The above probability mass function could be difficult to implement in practice due to the combinatorial subset calculation. Under model~\eqref{intensity}, we can change the order of summation to derive the total removal effect according to the linear decomposition of the direct removal effect in \eqref{dre_R}.
\begin{align*}
   \mathrm{att}_{ t_{i^\star}}^{\mathrm{(total)}}(R\mid D) =&\sum_{R\subseteq R' \subseteq\Omega}\mathrm{att}_{ t_{i^\star}}^{\mathrm{(direct)}}(R'\mid D)\mathbbm{P}(R^\diamond=R'\mid D)\\
   =&\sum_{R\subseteq R' \subseteq\Omega}\sum_{(t_i,e_i)\in R'}\mathrm{att}_{ t_{i^\star}}^{\mathrm{(direct)}}(\{(t_i,e_i)\}\mid D)\mathbbm{P}(R^\diamond=R'\mid D)\\
   =&\sum_{(t_i,e_i)\in \Omega}\mathrm{att}_{ t_{i^\star}}^{\mathrm{(direct)}}(\{(t_i,e_i)\}\mid D)\mathbbm{P}((t_i,e_i)\in R^\diamond\mid D).
\end{align*}
This result implies another iterative algorithm, which is simulation-free and more efficient. The basic idea is to redistribute the obtained scores from the direct removal effect. It adopts an intuitive backpropagation way of scoring as summarized in Algorithm~\ref{algo2'}, and the backpropagation, introduced by \cite{rumelhart1986learning}, has been gaining popularity in artificial neural networks. 

\begin{algorithm}[H]
  \caption{Total removal effect: backpropagation}
  \begin{algorithmic} 
\State Input: A path $D$, a specified conversion timestamp $ t_{i^\star}$ ($i^\star>1$) and a nonempty removal set $R\subseteq F_{t^\star}(D)$. Model parameters $\bm{\mu},\bm{\alpha}_1,\dots,\bm{\alpha}_q$.
\State Initialize: $y_i = \mathrm{att}_{ t_{i^\star}}^{\mathrm{(direct)}}(\{(t_i,e_i)\}\mid D)$ for $i_{\min} (R)\le i\le i^\star-1$.
      \For {$i$ in $\{i> i_{\min} (R):(t_i,e_i)\in F^{\mathcal{E}_\mathrm{c}}_{t_{i^\star}}(D\setminus R)\}$ (descending order)}
      \For {$i'$ in $\{1\le i'\le i-1:(e_{i'}\rightarrow e_i)\in \mathcal{A}\}$ }
      \State $y_{i'} = y_{i'} + y_{i}\cdot \Big(1-\frac{\lambda_{e_i}(t_i\mid \mathcal{H}_{t_i}^{D\setminus \{(t_{i'},e_{i'})\}})}{\lambda_{e_i}(t_i\mid \mathcal{H}_{t_i}^D)}\Big)$.
    \EndFor
    \EndFor
\State    
Return: $\mathrm{att}_{ t_{i^\star}}^{\mathrm{(total)}}(R\mid D)=\sum_{(t_i,e_i)\in R}y_i$.
 \end{algorithmic} 
  \label{algo2'}
\end{algorithm}

\begin{figure}[H]
\centering
\includegraphics[width=15.2cm]{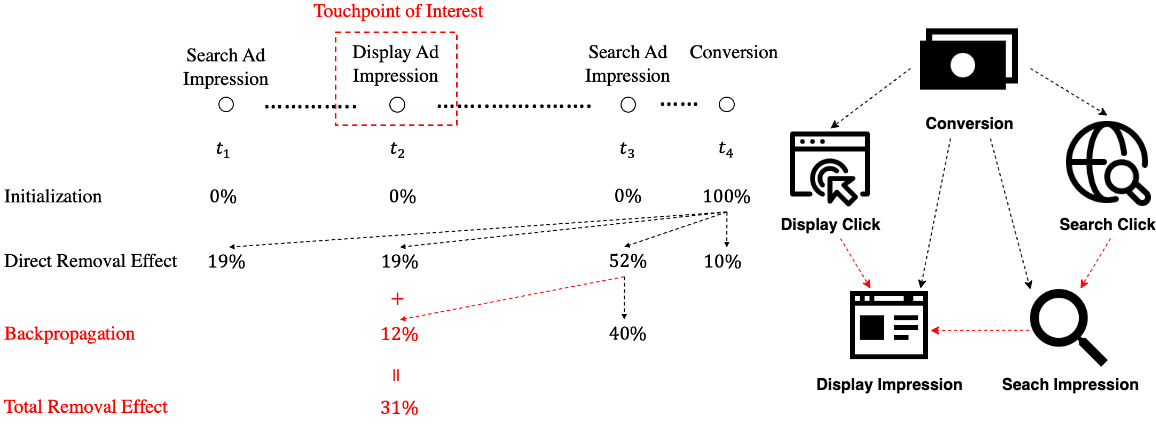}
\caption{For the example path containing three touchpoints, the total removal effect of the display ad impression can be calculated by the score backpropagation. It consists of two parts, the direct removal effect of itself and its share in the direct removal effect of the subsequent search ad impression. The graph on the right describes how the score flows between event types, with black and red arrows representing the direct removal effect attribution and score backpropagation, respectively. }\label{backprop}
\end{figure}

As shown by the example path in Figure~\ref{backprop}, instead of going forward along the path, the score flows backward from the last touchpoint to the previous touchpoints. This flow follows the reverse graph of the Granger causality graph, whose arrows point from an event type to its Granger causality parents.

\subsection{Remark on the Two Scoring Methods}
In this subsection, we discuss the difference between the proposed two attribution scoring methods and their implementation in an extreme case.

The direct removal effect regards the contributions from each touchpoint as individual components. While the total removal effect of a specified subset of touchpoints is the cumulative influence resulting from its removal along the path. The direct removal effect regards the conversion event as the only response, while the whole vector/subprocess of customer-initiated event types is the response for the total removal effect. Intuitively, on the score-flow graph in Figure~\ref{backprop}, the direct removal effect lets the score directly come out from the conversion node, and the total removal effect is based on the continual flow of the score until it reaches a specified node or a node with no parents.

These two scores differ in the context of additivity. Suppose $R_1$ and $R_2$ are two mutually exclusive subsets of the truncated path $F_{t^\star}(D)$. The direct removal effect under model~\eqref{intensity} has the additive property implied by the definition that $$\mathrm{att}_{ t_{i^\star}}^{\mathrm{(direct)}}(R_1\cup R_2\mid D) = \mathrm{att}_{ t_{i^\star}}^{\mathrm{(direct)}}(R_1\mid D) +\mathrm{att}_{ t_{i^\star}}^{\mathrm{(direct)}}( R_2\mid D).$$ However, for the total removal effect, we let the score flow between the two sets during backpropagation wherever there is Granger causality between their event types. Hence the total removal effect is subadditive and thus not incremental
$$\mathrm{att}_{ t_{i^\star}}^{\mathrm{(total)}}(R_1\cup R_2\mid D) \le \mathrm{att}_{ t_{i^\star}}^{\mathrm{(total)}}(R_1\mid D) +\mathrm{att}_{ t_{i^\star}}^{\mathrm{(total)}}( R_2\mid D).$$
As a result, to obtain the total removal effect of a channel, we should select the removal set as the set of all the corresponding individual touchpoints rather than take the sum of touchpoint-wise scores.

\begin{figure}[H]
\centering
\includegraphics[width=12cm]{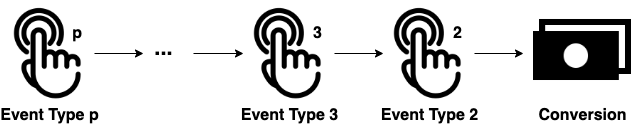}
\caption{An extreme case where the Granger causality graph is line-shaped.}\label{graph_line}
\end{figure}
We consider an extreme case illustrated by the Granger causality graph in Figure~\ref{graph_line}. In practice, we believe that touchpoints in each channel Granger-cause conversion separately, and hence the conversion node has multiple parents on the Granger causality graph like Figure~\ref{simgraph_hawkes}. But here we use this example to justify the proposed two attributions scores. On the graph, conversion is the event type labeled as $1$, and all the other event types $2,\dots,p$ are touchpoints, where the touchpoint type labeled as $p$ is the only firm-initiated event type. Suppose there is no baseline intensity for any customer-initiated event type, namely $\mu_e=0$ for $e=1,\dots,p-1$. Then a type-$(e+1)$ event will be the only reason for the occurrence of a type-$e$ event. Suppose we observe a path $D$ with the same pattern as Figure~\ref{graph_line}. That is,
\[D=\{(t_1,p),\dots,(t_{p-2},3),(t_{p-1},2),(t_p,1)\}.\]
If we look at the direct removal effect of each touchpoint, we get \[\mathrm{att}_{ t_{p}}^{\mathrm{(direct)}}(\{(t_i,e_i)\}\mid D)=\begin{cases}100\%, &i=p-1;\\
0\%,& i=1,\dots,p-2,
\end{cases}\]
where $e_i = p-i+1$ for $i=1,\dots,p-1$.
The direct removal effect ignores the importance of touchpoints in the early stage since they cannot instantly trigger a conversion. On the other hand, the total removal effect gives \[\mathrm{att}_{ t_{p}}^{\mathrm{(total)}}(\{(t_i,e_i)\}\mid D)=100\%,\quad i=1,\dots,p-1.
\] It seems that the total removal effect can over-allocate since there is only $1$ conversion on path $D$. This issue was also pointed out by \cite{singal2022shapley} as a drawback of attribution methods based on the removal effect. However, it does not mean the scores are wrong. This extreme over-allocation problem implies that it might be inappropriate to calculate such an attribution score for each touchpoint. In Figure~\ref{graph_line}, it is likely that event types $2$ to $p$ all belong to the same channel. A detailed example can be the case where email sent, email open, and email click are the touchpoints, and conversion can be only triggered by email click (invite-only purchase through email link). In this case, we can regard event types $2$ to $p$ as an entirety using the removal set $R = \{(t_1,p),\dots,(t_{p-2},3),(t_{p-1},2)\}$. Then we have
\[\mathrm{att}_{ t_{p}}^{\mathrm{(direct)}}(R\mid D)=\mathrm{att}_{ t_{p}}^{\mathrm{(total)}}(R\mid D)=100\%.
\]
If event types $2$ to $p$ are actually from different channels, it means the graph is not likely true, or there is no need to distinguish these channels from each other since they are highly dependent.

To summarize, the direct removal effect serves as an explanatory score by allocating the credit of a conversion event to each touchpoint according to its incremental influence. Both proposed methods can give channel-level scores, but the total removal effect provides a marginal point of view for conversion. In other words, it better applies to cases where the leave-one-channel-out loss of conversion is the desired quantity. 

\section{Estimation}
In this section, we develop a regularized estimator for our proposed model. We design an computationally efficient alternating direction method of multipliers (ADMM) algorithm to solve the corresponding optimization problem. We also derive the graphical model selection consistency and the rates of convergence rate of model estimates and attribution scores in the asymptotic regime. 

We introduce some necessary notations to facilitate the presentation. For any $K$-dimensional vector $\bm{x}=(x_1,\dots,x_K)^\top\in\mathbb{R}^{K}$, where $K\ge 1$ is an arbitrary integer, let $\|\bm{x}\|_1:=\sum_{k=1}^K |x_k|$, $\|\bm{x}\|_2:=\sqrt{\sum_{k=1}^K x_k^2}$, and $\|\bm{x}\|_\infty:=\max_{1\le k \le K} |x_k|$ denote its $L_1$-norm, $L_2$-norm, and $L_\infty$-norm respectively.
Let $\bm{x}_{+}$ denote the non-negative part of $\bm{x}$ that is defined by $\bm{x}_{+}:= (\max(0,x_1),\dots,\max(0,x_K))^\top$. For a matrix $M=(M_{ij})_{i,j}$, the matrix $L_1$-norm is defined as $\big\|M\big\|_{1,\infty}:=\max_i(\sum_j \big|M_{ij}\big|)$.
\subsection{Model Learning}
Suppose there are $n$ paths in total, $D_{1},\dots,D_{n}$, where $D_{j}$ is the $j$-th path on time interval $[0,T_{j}]$. Suppose the kernel functions $\psi_{e'e}(\cdot),\ e'=1,\dots,p,\ e = 1,\dots,q,$ are known. Let $\bm{\mu} = (\mu_1,\dots,\mu_q)^\top\in \mathbb{R}_{\ge 0 }^q$ be the vector of baseline intensities of $\mathbf{N}_{\mathcal{E}_\mathrm{c}}(t)$ and $\bm{\alpha}_e = (\alpha_{1e},\dots,\alpha_{pe})^\top\in \mathbb{R}_{\ge 0 }^p$ be the $e$-th column of $A$ for $e = 1,\dots,q$. Learning the model turns into an optimization problem:
$$\min_{\bm{\mu},\bm{\alpha}_1,\dots,\bm{\alpha}_q\in \mathbb{R}_{\ge 0 }^p}\frac1n\sum_{j=1}^n\Phi(D_{j};\bm{\mu},\bm{\alpha}_1,\dots,\bm{\alpha}_q),$$where $\Phi(D_{j};\bm{\mu},\bm{\alpha}_1,\dots,\bm{\alpha}_q)$ is the loss function given the $j$-th path $D_j$. Consider a least-squares functional:
    \[\Phi(D_{j};\bm{\mu},\bm{\alpha}_1,\dots,\bm{\alpha}_q) = \sum_{e=1}^q\left\{\frac12\int_0^{T_{j}} [\lambda_e(t\mid \mathcal{H}_{t}^{D_{j}})]^2 dt-\int_0^{T_{j}} \lambda_e(t\mid \mathcal{H}_{t}^{D_{j}}) dN^{D_{j}}_e(t)\right\}.\]
Compared with the negative log-likelihood, the least-squares functional enjoys better computational efficiency. Its equivalent form was proposed for estimating the additive risk model by \cite{lin1994semiparametric}. It was also adopted by \cite{hansen2015lasso} and \cite{bacry2020sparse} for point process models.
    
Assume that the edge set is sparse, we add the sparsity constraints on the coefficients as follows: \begin{equation}\label{joint}
\min_{\bm{\mu},\bm{\alpha}_1,\dots,\bm{\alpha}_q \in \mathbb{R}_{\ge 0 }^p}\frac1n\sum_{j=1}^n \Phi(D_{j};\bm{\mu},\bm{\alpha}_1,\dots,\bm{\alpha}_q) + \sum_{e=1}^q\gamma_e \| \bm{\alpha}_e\|_1,
\end{equation}
where $\gamma_1,\dots,\gamma_q \ge 0$ are the regularization parameters to control the individual sparsity of coefficient vectors $\bm{\alpha}_1,\dots,\bm{\alpha}_q$. It is worth pointing out that $\{\mu_e,\bm{\alpha}_e\}_{e=1}^q$ are separable in the objective function \eqref{joint}. Thus, the optimization problem can be decomposed into node-wise model learning. For each node $e\in\mathcal{E}_\mathrm{c}$, learning its parent nodes yields 
\begin{equation}\label{nodewise}
    \min_{\mu_e\ge 0 ,\bm{\alpha}_e\in \mathbb{R}_{\ge 0 }^p}\frac1n\sum_{j=1}^n\phi_e(D_{j};\mu_e,\bm{\alpha}_e) + \gamma_e \| \bm{\alpha}_e\|_1,
\end{equation}
where
\[
\phi_e(D_{j};\mu_e,\bm{\alpha}_e) =\frac12\int_0^{T_{j}} [\lambda_e(t\mid \mathcal{H}_{t}^{D_{j}})]^2dt-\int_0^{T_{j}}\lambda_e(t\mid \mathcal{H}_t^{D_{j}}) dN^{D_{j}}_e(t).\]

 In the following context, we write $\bm{\theta} =(\theta_0,\theta_1,\dots,\theta_p)^\top= (\mu_e,\bm{\alpha}_e^\top)^\top\in\mathbb{R}_{\ge 0}^{p+1}$. Let $\bm{X}_{j}(t)=(X_{j,0}(t),X_{j,1}(t),\dots,X_{j,p}(t))^\top$ with \[X_{j,0}(t) = 1, \quad X_{j,k}(t)=\int_0^t \psi_{ke}(t-u) dN^{D_{j}}_k(u),\ k=1,\dots,p.\]
Now the conditional intensity can be written as $\lambda_e(t\mid \mathcal{H}_t^{D_{j}}) = \bm{\theta}^\top\bm{X}_{j}(t)$.
Let $V=(V_{kk'})_{k,k'=0}^p\in\mathbb{R}^{(p+1)\times (p+1)}$ and $\bm{b}=(b_0,\cdots,b_p)^\top\in\mathbb{R}^{p+1}$, where for $k=0,\dots,p$,
\[V_{kk'}= \frac1n\sum_{j=1}^n \int_0^{T_{j}}  X_{j,k}(t)X_{j,k'}(t)dt, \quad b_k= \frac1n\sum_{j=1}^n \int_0^{T_{j}}  X_{j,k}(t)dN_e^{D_{j}}(t).\]
Then the regularized solution satisfies
\begin{equation}\label{quadratic}
    \hat{\bm{\theta}}=\underset{\bm{\theta}\in\mathbb{R}_{\ge 0}^{p+1}}{\arg\min}\ \frac12\bm{\theta}^\top V\bm{\theta}-\bm{b}^\top\bm{\theta}+\gamma_e\|\bm{\alpha}_e\|_1.
\end{equation}

The above problem is equivalent to a linearly constrained one
\[\min_{\bm{\theta}\in\mathbb{R}_{\ge 0}^{p+1}, \bm{\alpha}_e=\bm{\alpha}_e'}\ \frac12\bm{\theta}^\top V\bm{\theta}-\bm{b}^\top\bm{\theta}+\gamma_e\|\bm{\alpha}_e'\|_1.\]
Apply the alternating direction method of multipliers (ADMM), where the corresponding augmented Lagrangian function is
\[\mathcal{L}_\eta(\bm{\theta},\bm{\alpha}_e',\bm{\omega})=\frac12\bm{\theta}^\top V\bm{\theta}-\bm{b}^\top\bm{\theta} + \gamma_e \|\bm{\alpha}_e'\|_1+ \bm{\omega}^\top(\bm{\alpha}_e- \bm{\alpha}_e')+\frac12\eta\|\bm{\alpha}_e- \bm{\alpha}_e'\|_2^2.\]
The learning algorithm is shown in Algorithm~\ref{algo:ADMM}.
\begin{algorithm}[H]
  \caption{Graphical point process learning by ADMM}
  \begin{algorithmic} 
\State Input: Paths $D_1,\dots,D_n$ and the regularization parameter $\gamma_e$. 
\State Pre-compute: $V,\ \bm{b}$.
\State Initialize: Set $\eta>0$ and proper initial values of $\mu_e$, $\bm{\alpha}_e$, $\bm{\alpha}_e'$, and $\bm{\omega}$.

    \While {not converge}\vspace{0.5em}
      \State $\begin{pmatrix}\mu_e\\\bm{\alpha}_e\end{pmatrix} = \left[\left(V+ \begin{pmatrix}0&\\&\eta I_p\end{pmatrix}\right)^{-1}\left(\bm{b}+\begin{pmatrix}0\\\eta\bm{\alpha}_e'-\bm{\omega}\end{pmatrix}\right)\right]_{+}$.
      \vspace{0.5em}
      \State\; $\bm{\alpha}_e' =(\bm{\alpha}_e+\eta^{-1}\bm{\omega}-\eta^{-1}\gamma_e\mathbbm{1}_p)_{+}$.
      \State\; $\bm{\omega} = \bm{\omega}+\eta(\bm{\alpha}_e- \bm{\alpha}_e')$.
    \EndWhile
    \State Return: $\bm{\theta}=(\mu_e,\bm{\alpha}_e^\top)^\top$.
 \end{algorithmic}
 \label{algo:ADMM}
\end{algorithm}

\subsection{Asymptotic Properties}
In this subsection, we derive the rate of convergence of the proposed estimator. The customer-initiated part of our model~\eqref{intensity} is similar to Hawkes process \citep{hawkes1971spectra}. But unlike the typical multivariate Hawkes process, model~\eqref{intensity} is not guaranteed a stationary point process. As a result, the asymptotic results are derived under the assumption that the sample size $n\to \infty$, instead of the length of observation $T \to \infty$ \citep{guo2018consistency,yu2020hawkesian}. In survival analysis, \cite{lin1994semiparametric} studied the additive risk model, and \cite{lin2013high} established the consistency of the corresponding $L_1$-regularized estimator. Combining these works, we will show that under certain conditions, including irrepresentability, our proposed estimator is consistent in both the classical fixed $p$ setting and the sparse high-dimensional setting, where $\log(p)$ is comparable to the sample size $n$.

\begin{assumption}(I.I.D.) The process of the customer-initiated event types $\mathbf{N}_{\mathcal{E}_\mathrm{c}}^{D_{j}}(t)$, $j=1,\dots,n$ are independent and follow model~\eqref{intensity}. 
\end{assumption}
The convergence properties of the estimator rely on the identical distribution of the observations. We do not need to assume the external (firm-initiated) events are \emph{I.I.D.} across paths.
They may vary from one path to another, but model~\eqref{intensity} should be true for each individual.
\begin{assumption}(Bounded input)
There exist constants $\overline{T}$ and $\overline{X}$ such that $T_{j}<\overline{T}$ and $\sup_{t\in (0,T_{j})} \big\|\bm{X}_{j}(t)\big\|_\infty<\overline{X}$ a.s.
for $j = 1,\dots,n$.
\end{assumption}
Define the active set $S =\{0 \}\cup \{e'\in \mathcal{E}:\alpha_{e'e}> 0 \}$ and its complement $S^{\mathsf{c}} = \{e'\in \mathcal{E}:\alpha_{e'e}= 0 \}$. We use $s = \mathrm{Card}(S)$ to denote the cardinality of the active set $S$. Let $G=(G_{kk'})_{k,k'=0}^p=\mathbbm{E} V$ be the population version of $V$. Assume the sub-matrix $G_{SS}=(G_{kk'})_{k,k'\in S}$ is non-singular and define $\kappa := \big\|G_{SS}^{-1}\big\|_{1,\infty}$.

\begin{assumption}(Irrepresentability) There exists a constant $\xi\in (0,1)$ such that
\[\big\|G_{S^{\mathsf{c}}S}G_{SS}^{-1}\big\|_{1,\infty}<1-\xi.\]
\end{assumption}
This condition is adapted from Condition 3 in \cite{lin2013high}, which is a generalization of Condition (15) in \cite{wainwright2009sharp} for linear regression with LASSO.
 
Now we establish the rate of convergence and the model selection consistency of the regularized estimator. By abuse of notation, let $\bm{\theta}$ denote the true value. We consider two scenarios, where the number of event types $p$ is fixed or $p$ diverges while the active set $S$ is sparse in the sense that $s$ is bounded from $p$.
\begin{theorem} Under Assumptions 1-3, there exist $C_1>0$ and $C_2>0$ such that the regularized estimator $\hat{\bm{\theta}}$ in \eqref{quadratic} satisfies the following properties:
\begin{enumerate}
    \item [(i)] If $p$ is fixed, then for any constant $0<\nu<1$, when $\gamma_e$ is chosen properly, and $n$ is sufficiently large, with probability at least $1-2(p+1)(p+2)\exp(-n^\nu)$,
\begin{itemize}
    \item  (Edge selection) $\hat{\bm{\theta}}_{S^{\mathsf{c}}}=0$
    \item  ($L_\infty$-error) $\big\|\hat{\bm{\theta}}-\bm{\theta}\big\|_\infty\le 10\xi^{-1}\kappa C_1^{-\frac12}C_2 n^{-\frac{1-\nu}{2}}$;
\end{itemize}
    \item  [(ii)] If $p$ diverges, then for any constant $\zeta>2$, when $\gamma_e$ is chosen properly, and $n$ is sufficiently large, with probability at least $1-3/(p+1)^{\zeta-2}$,
\begin{itemize}
    \item  (Edge selection) $\hat{\bm{\theta}}_{S^{\mathsf{c}}}=0$
    \item  ($L_\infty$-error) $\big\|\hat{\bm{\theta}}-\bm{\theta}\big\|_\infty\le 10\xi^{-1}\kappa C_1^{-\frac12}C_2\sqrt{\frac{\zeta\log(p+1)}{n}}$.
\end{itemize}
\end{enumerate}
\label{consistency_theta}
\end{theorem}
In the above Theorem, we do not specify the choices of $\gamma_e$ and $n$ for ease of presentation. The choices in detail can be found in the supplementary materials.

Then we move on to the rate of convergence of attribution scores. For a positive path $D$ with a conversion at $t=t_{i^\star}$, we would like to analyze the direct removal effect of a subset $R\subseteq F_{t^\star}(D)$. In practice, for Equation \eqref{dre}, we can only obtain the estimate of the conditional intensities. As a result, for model~\eqref{intensity}, we use the estimated direct removal effect given by
$$ \widehat{\mathrm{att}}_{ t_{i^\star}}^{\mathrm{(direct)}}(R\mid D) :=\frac{\sum_{(t_i,e_i)\in R}\hat{\alpha}_{e_i1} \psi_{e_i1}( t_{i^\star}-t_i)}{ \hat{\mu}_1+\sum_{i < i^\star}\hat{\alpha}_{e_i1}\psi_{e_i1}( t_{i^\star}-t_{i})}.$$
Let $r = \mathrm{Card}(R)$ denote the cardinality of the removal set $R$. For the kernel functions, suppose there exists $\overline{\psi}_1>0$ such that $\max_{1\le e\le p}\sup_{t>0}|\psi_{e1}(t)|<\overline{\psi}_1$. 
\begin{theorem} Given a path $D$ with a conversion at $t=t_{i^\star},\ i^\star>1$, under the above assumptions for $e=1$, there exists $C_3>0$ dependent on $\overline{\psi}_1$ and $D$ such that the estimated direct removal effect of the removal set $R\subseteq F_{t^\star}(D)$ satisfies either of the following condition: 
\begin{enumerate}
    \item [(i)] If $p$ is fixed and $n$ is sufficiently large, then for any constant $0<\nu<1$, with probability at least $1-2(p+1)(p+2)\exp(-n^\nu)$, $$\left|\widehat{\mathrm{att}}_{ t_{i^\star}}^{\mathrm{(direct)}}(R\mid D)-\mathrm{att}_{ t_{i^\star}}^{\mathrm{(direct)}}(R\mid D)\right|\le 10\xi^{-1}\kappa C_1^{-\frac12}C_2C_3 r n^{-\frac{1-\nu}{2}};$$ 
    \item [(ii)] If $p$ diverges and $n$ is sufficiently large, then for any constant $\zeta>2$, with probability at least $1-3/(p+1)^{\zeta-2}$,
    $$\left|\widehat{\mathrm{att}}_{ t_{i^\star}}^{\mathrm{(direct)}}(R\mid D)-\mathrm{att}_{ t_{i^\star}}^{\mathrm{(direct)}}(R\mid D)\right|\le   10\xi^{-1}\kappa C_1^{-\frac12}C_2C_3 r\sqrt{\frac{\zeta\log(p+1)}{n}}.$$
\end{enumerate}
\label{consistency_dre}
\end{theorem}
The constants $\xi$, $\kappa$, $C_1$, and $C_2$ come from Theorem~\ref{consistency_theta} corresponding to $e = 1$. Based on the analysis of the estimated direct removal effect, we then provide an error bound of the total removal effect in estimation. We refer to the estimated total removal effect as
$$ \widehat{\mathrm{att}}_{ t_{i^\star}}^{\mathrm{(total)}}(R\mid D) :=\mathbbm{E}[\widehat{\mathrm{att}}_{ t_{i^\star}}^{\mathrm{(direct)}}(\widehat{R^\diamond}\mid D)\mid D],$$
where $\widehat{R^\diamond}\supseteq R$ is the actual removal set obtained by the thinning operation in Algorithm~\ref{algo2} using estimated thinning probabilities. Besides $\overline{\psi}_1$, for each $e=2,\dots,q$, suppose there exists $\overline{\psi}_e>0$ such that $\max_{1\le e'\le p}\sup_{t>0}|\psi_{e'e}(t)|<\overline{\psi}_e$.
\begin{theorem} Given a path $D$ with a conversion at $t=t_{i^\star},\ i^\star>1$ and a removal set $R\subseteq F_{t^\star}(D)$, under the above assumptions for $e=1,\dots,q$, there exists $C_4>0$ dependent on $\overline{\psi}_1,\dots,\overline{\psi}_q$, $D$, and $R$ such that the estimated total removal effect of $R$ satisfies either of the following condition:
\begin{enumerate}
    \item [(i)] If $p$ is fixed and $n$ is sufficiently large, then for any constant $0<\nu<1$, with probability at least $1-2p(p+1)(p+2)\exp(-n^\nu)$, $$\left|\widehat{\mathrm{att}}_{ t_{i^\star}}^{\mathrm{(total)}}(R\mid D)-\mathrm{att}_{ t_{i^\star}}^{\mathrm{(total)}}(R\mid D)\right|\le  C_4  n^{-\frac{1-\nu}{2}};$$ 
    \item [(ii)] If $p$ diverges and $n$ is sufficiently large, then for any constant $\zeta>3$, with probability at least $1-3/(p+1)^{\zeta-3}$,
    $$\left|\widehat{\mathrm{att}}_{ t_{i^\star}}^{\mathrm{(total)}}(R\mid D)-\mathrm{att}_{ t_{i^\star}}^{\mathrm{(total)}}(R\mid D)\right|\le  C_4\sqrt{\frac{\zeta\log(p+1)}{n}}.$$
\end{enumerate}
\label{consistency_tre}
\end{theorem}

\section{Simulation Study}
In this section, we carry out two simulation experiments to examine the performance of our graphical attribution methods. In the first part, we validate the proposed graphical attribution methods using data simulated from the multivariate Hawkes process. In the second part, we compare the proposed methods with the commonly used attribution models using data simulated from a modified version of the Digital Advertising System Simulation (DASS) developed by Google Inc. The simulated data includes online customer browsing behavior and injected advertising events that impact this customer behavior.

We first explain some channel-level metrics for attribution methods. In general, suppose there are $Z$ channels, labeled as $z=1,\dots,Z$. Inspired by the previous literature \cite{anderl2016mapping, li2014attributing}, we calculate the proportion of channel-level conversion count (proportion of CCC) for each channel $z$, which is \[\mathsf{p}_z := \frac{\mathrm{CCC}_z}{\sum_{z=1}^Z\mathrm{CCC}_z}\quad \text{with}\ \mathrm{CCC}_z=\sum_{j=1}^n N_1^{D_j}((0,T_j])-\sum_{j=1}^n N_1^{D_j^\text{($z$-off)}}((0,T_j]),\]
where $D_j^\text{($z$-off)}$ is a path following the same distribution as $D_j$ other than having no touchpoints of channel $z$. $\mathrm{CCC}_z$ is interpreted as the number of conversions resulting from channel $z$.
The first term of $\mathrm{CCC}_z$ is the number of conversions out of $n$ paths, and the second term is the number of conversions out of $n$ paths when channel $z$ is turned off. To obtain this quantity for synthetic data, we can disable all the related touchpoint types and run the simulator again with the same seed. Then the decrease in total conversions is the desired value.

Let $\mathcal{C}_z\subseteq \mathcal{E}$ denote the set of touchpoint types belonging to channel $z$, for $z=1,\dots,Z$. Then the corresponding removal set with respect to the conversion at $t$ in the path $D$ is $F^{\mathcal{C}_z}_{t}(D)=\{(t_i,e_i)\in D:t_i<t,\ e_i\in\mathcal{C}_z\}$. We use the proportion of channel-level aggregated score (proportion of CAS) to estimate the proportion of CCC for channel $z$, which is
\[\hat{\mathsf{p}}_z := \frac{\mathrm{CAS}_z}{\sum_{z=1}^Z\mathrm{CAS}_z}\quad \text{with}\ \mathrm{CAS}_z=\sum_{j=1}^n\sum_{i\le m_j:\ e_i^j=1} \widehat{\mathrm{att}}_{ t_{i}^j}(F^{\mathcal{C}_z}_{t_{i}^j}(D_{j})\mid D_{j}).\]
Recall that the attribution score is an incremental component or marginal loss with respect to conversion. Its aggregated version, CAS, is the overall decrease in conversion counts for each channel compared with the total number of conversions and thus can be used to estimate CCC.
Let $\bm{\mathsf{p}}=(\mathsf{p}_1,\dots,\mathsf{p}_Z)^\top$ denote the vector of proportions of CCC and $\hat{\bm{\mathsf{p}}}=(\hat{\mathsf{p}}_1,\dots,\hat{\mathsf{p}}_Z)^\top$ denote the vector of the proportions of CAS. We adopt the KL divergence $D_\mathrm{KL}(\cdot\parallel\cdot)$ and the Hellinger distance $H(\cdot,\cdot)$ to measure the estimation accuracy, where
\begin{align*}
   D_\mathrm{KL}(\bm{\mathsf{p}}\parallel \hat{\bm{\mathsf{p}}})&:=\sum_{z=1}^Z\mathsf{p}_z \log\left(\frac{\mathsf{p}_z}{\hat{\mathsf{p}}_z}\right)\\
    H(\bm{\mathsf{p}}, \hat{\bm{\mathsf{p}}})&:=\sqrt{\frac{1}{2}\sum_{z=1}^Z\left(\sqrt{\hat{\mathsf{p}}_z}-\sqrt{\mathsf{p}_z}\right)^2}.
\end{align*}
\subsection{Simulation Based on Hawkes Process}
In this subsection, we verify our attribution methods using a dataset simulated from the multivariate Hawkes process \citep{hawkes1971spectra}. 

Referring to Equation \eqref{intensity}, our model reduces to a multivariate Hawkes process if $N_e(t)$ is a Poisson process for each $e\in\mathcal{E}_\mathrm{f}$. In other words, the multivariate Hawkes process is nested in our model. Therefore, we simulate a data set using a multivariate Hawkes process according to Figure~\ref{simgraph_hawkes}, which involves $Z=2$ channels, display and search, and $4$ types of touchpoints: display impression, display click, search impression, and search click. Display impression is regarded as a firm-initiated event type, following a Poisson process distribution with a rate of $0.02$. A total of $n=10,000$ paths are simulated with $T_j=365$ days for $j=1,\dots,n$. We take $\psi_{e'e} = \frac{1}{10}\exp(-\frac{t}{10})\cdot\mathbbm{1}_{\{t>0\}}$ for each possible pair of connected nodes. The baseline intensities of search impression and conversion are set as $0.02$ and $1\times 10^{-4}$, and the two click touchpoint types have zero baselines.

\begin{table}
\centering
{\begin{tabular}{c|cccc}
\hline
\diagbox[linewidth=0.4pt, width=\dimexpr \textwidth/4+2\tabcolsep\relax, height=0.8cm]{From}{To} &Display click & Search impression & Search click & Conversion \\ \hline
Display impression                          & 0.08                                                         & 0.08              & 0            & 0.01       \\
Display click                               & 0                                                            & 0                 & 0            & 0.08       \\
Search impression                           & 0                                                            & 0                 & 0.08         & 0.02       \\
Search click                                & 0                                                            & 0                 & 0            & 0.1        \\
Conversion                                  & 0                                                            & 0                 & 0            & 0      \\
\hline
\end{tabular}}
\caption{Granger causality coefficients of the simulated data.}
\label{para_hawkes}
\end{table}

\begin{table}[H]
\centering
\begin{tabular}{c|cc}
\hline
\makebox[2cm][l]{Channel:\hfill}$\mathsf{p}_z$ & TRE & DRE \\ \hline
\makebox[1.5cm][l]{Display:\hfill} $0.3799$    & $0.3782\ (0.0104)$   & $0.3491\ (0.0112)$  \\
\makebox[1.5cm][l]{Search:\hfill} $0.6201$   & $0.6218\ (0.0104)$  & $0.6509\ (0.0112)$ \\ \hline
KL divergence     & $0.0002\ (0.0002)$   & $0.0023\ (0.0015)$  \\
Hellinger distance & $0.0064\ (0.0042)$  & $0.0226\ (0.0083)$ \\
\hline
\end{tabular}%
\caption{Comparison of the proportions of CAS between total removal effect and direct removal effect. Reported numbers are averages over $100$ independent runs, with standard errors given in parentheses. The first column lists the proportions of channel-level conversion counts (CCC) as the ground truth.}
\label{simtb_hawkes}
\end{table}

The Granger causality coefficients are given in Table~\ref{para_hawkes}. The simulation results are summarized in Table~\ref{simtb_hawkes}, which are calculated over $100$ independent runs. As shown in Table~\ref{simtb_hawkes}, our graphical TRE attribution method, which takes into account the Granger causality among different types of touchpoints is accurate in estimating the true removal effects of both search and display channels. The proportions of CAS for both channels calculated by TRE are very close to the ground truth. Also, the KL divergence and the Hellinger distance of TRE are very small. This is not surprising since Theorem~\ref{thinning} guarantees that $D_j\setminus(F^{\mathcal{C}_z}_{T_j}(D_{j}))^\diamond$ and $D_j^\text{($z$-off)}$ are identically distributed. In contrast, the graphical DRE method tends to underestimate the contribution of the display channel because it ignores the exciting effect of the display impression on the search impression. 

\subsection{Simulation Based on DASS}
Next, we compare our graphical attribution methods with existing methods, including DNAMTA \citep{li2018deep}, logistic regression, Markov model \citep{anderl2016mapping}, as well as the rule-based methods including last-touch, first-touch, linear, time-decay, and U-shaped. Table~\ref{simtb_benchmark} lists the description of the models under comparison. Among them, the Markov model provides channel-level scores directly and the others provide path-level scores that can be aggregated to the channel level.
    
\begin{table}[ht]
\resizebox{\textwidth}{!}{\begin{tabular}{llm{15.2cm}}
\hline
\textbf{Method}     &\textbf{Type}                             &  \textbf{Scoring Description}    \\
\hline
TRE                            &Data-driven& Total removal effect of the graphical point process model.                                                                                                                                             \\
DRE                             &Data-driven& Direct removal effect of the graphical point process model.                                                                                                                                                   \\
DNAMTA                        &Data-driven& An incremental score derived from the conversion probability of Deep Neural Net With Attention multi-touch attribution model developed in \cite{li2018deep}.                                                                                                        \\
Logistic                    &Data-driven& An incremental score derived from the conversion probability of logistic regression.                                                                                                                                              \\
Markov                        &Data-driven& Removal effect of Markov model developed in \cite{anderl2016mapping}.                                                                                              \\

Last       & Rule-based& Last-touch attribution, assigning all credit to the touchpoint closest to the conversion.\\

First      & Rule-based& First-touch attribution, assigning all credit to the initial touchpoint on a path.                                                                                       \\

Linear         & Rule-based& Linear attribution, assigning equal credit to each touchpoint before the conversion.                                                                                                 \\

Decay & Rule-based& Time-decay attribution, where touchpoints closer to the conversion receive more credit than touchpoints that are farther away in time from the conversion.                                  \\

U-shaped       & Rule-based& U-shaped attribution, assigning $40\%$ of the credit to both the first touchpoint and the last touchpoint, with the other touchpoints splitting the remaining $20\%$ equally.
\\
\hline
\end{tabular}}
\caption{Description of attribution methods under comparison.}
\label{simtb_benchmark}
\end{table}
For model comparison, we simulate data from a modified version of the Digital Advertising System Simulation (DASS). DASS \citep{sapp2016dass}, developed by Google Inc., is a popular attribution simulator in the industry and its
effectiveness is well accepted by practitioners. ``It generates the data to
which observational models can be applied, as well as the ability to run virtual experiments with simulated customers to measure the actual incremental value of marketing for direct comparison" \citep{sapp2016dass}. We modified DASS to work with two desired features and call it DASS+. DASS simulates transitions between the browsing states of each customer without any clear regard for timestamps while DASS+ uses a transition matrix reflecting these browsing states in each minute. On the other hand, DASS has no explicit restriction on the number of advertisements that can be served. With DASS+, the number of ads is capped to a fixed amount, and the ads can be served in a pre-determined distribution across the simulation timeframe.

This synthetic data involves $Z=4$ channels, email, display, search, and social including $9$ types of touchpoints: email sent, email open, email click, display impression, display click, search impression, search click, social impression, and social click. With all channels turned on, we obtain $n=98,986$ valid paths out of $100,000$ customers. Among them, there are $62,287$ positive paths and $36,699$ negative paths. The simulation period is $90$ days for each path. For model learning, we use the timestamp of the very first event as the starting time ($t_1=0$) and use the timestamp of the final event on the path as the terminal time. 

\begin{figure}[H]
\centering
\includegraphics[width=8cm]{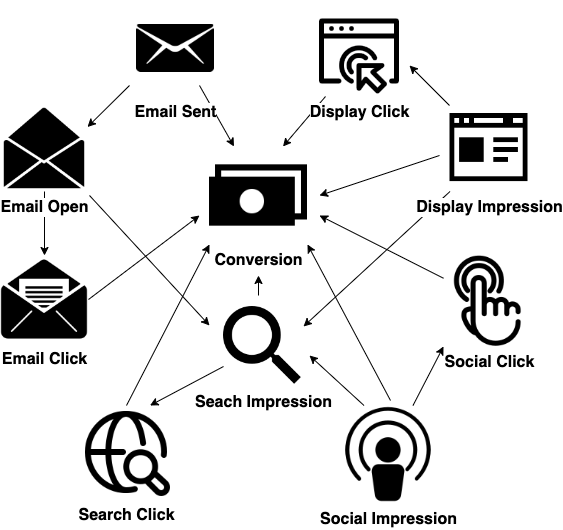}
\caption{The Granger causality graph for simulated data.}\label{simgraph}
\end{figure}

Figure~\ref{simgraph} shows the Granger causality graph learned by our proposed model for simulated data. Most types of touchpoints have exciting effects on conversion. Similar to the first simulation study, we confirm the intra-channel carry-over effects for every channel, even with more channels present. Also, we confirm the inter-channel spill-over effects from other channels to search impression.

\begin{table}[H]
\resizebox{\textwidth}{!}{%
\begin{tabular}{c|cccccccccc}
\hline
\makebox[1.8cm][l]{Channel:\hfill}$\mathsf{p}_z$ &
TRE &
DRE &
DNAMTA &
Logistic &
Markov &
Last &
First&
Linear &
Decay&
U-shaped\\ \hline
\multirow{2}{*}{\makebox[1.5cm][l]{Display:\hfill}$0.207$}   & $0.185$   & $0.137$   & $0.172$   & $0.275$   & $0.221$   & $0.371$   & $0.397$   & $0.380$   & $0.372$   & $0.383$   \\
                                                     & $(0.006)$ & $(0.006)$ & $(0.027)$ & $(0.004)$ & $(0.000)$ & $(0.002)$ & $(0.003)$ & $(0.001)$ & $(0.001)$ & $(0.001)$ \\
\multirow{2}{*}{\makebox[1.5cm][l]{Email:\hfill}$0.171$}  & $0.139$   & $0.101 $  & $0.130$   & $0.217$   & $0.356$   & $0.277$   & $0.293$   & $0.288$   & $0.282$   & $0.286$   \\
                                                 & $(0.008)$ & $(0.007)$ & $(0.019)$ & $(0.004)$ & $(0.000)$ & $(0.001)$ & $(0.002)$ & $(0.001)$ & $(0.001)$ & $(0.001)$ \\
\multirow{2}{*}{\makebox[1.5cm][l]{Search:\hfill}$0.485$}   & $0.574$   & $0.680$   & $0.579$   & $0.324$   & $0.206$   & $0.135$   & $0.078$   & $0.108$   & $0.127$   & $0.107 $  \\
                                                 & $(0.006)$ & $(0.007)$ & $(0.034)$ & $(0.003)$ & $(0.000)$ & $(0.001)$ & $(0.001)$ & $(0.000)$ & $(0.000)$ & $(0.001)$ \\
\multirow{2}{*}{\makebox[1.5cm][l]{Social:\hfill}$0.137$}   &$0.102$   & $0.082$   & $0.119$   & $0.183$   & $0.217$   & $0.217$  & $0.232$   & $0.224$   & $0.219$   & $0.224$   \\
                                                      & $(0.009)$ & $(0.008)$ & $(0.028)$ & $(0.004)$ & $(0.000)$ & $(0.002)$ & $(0.002)$ & $(0.001)$ & $(0.001)$ & $(0.001)$ \\ \hline
\multirow{2}{*}{KL divergence}    & $0.008$   & $0.011$   & $0.012$   & $0.024$   & $0.093$   & $0.154$   & $0.255$   & $0.194$   & $0.164$   & $0.196$   \\
& $(0.002)$ & $(0.001)$ & $(0.007)$ & $(0.002)$ & $(0.002)$ & $(0.004)$ & $(0.005)$ & $(0.004)$ & $(0.004)$ & $(0.004)$ \\
\multirow{2}{*}{Hellinger distance} & $0.067 $  & $0.141$   & $0.078$   & $0.117$   & $0.226$   & $0.277$   & $0.342$   & $0.306$   & $0.285$   & $0.307$   \\
& $(0.007)$ & $(0.007)$ & $(0.023)$ & $(0.005)$ & $(0.003)$ & $(0.004)$ & $(0.003)$ & $(0.003)$ & $(0.003)$ & $(0.003)$ \\
\hline
\end{tabular}%
}
\caption{Comparison of proportions of the channel-level aggregated score (CAS) for different methods. Reported numbers are averages over $10$ independent runs, with standard errors given in parentheses. The first column lists the proportions of channel-level conversion counts (CCC) as the ground truth.}
\label{simtb}
\end{table}

Table~\ref{simtb} summarizes the proportions of CAS, as well as the KL divergence and the Hellinger distance for different methods, which are calculated over $10$ independent runs. Our two graphical attribution methods, achieve the most accurate results. The rule-based methods (i.e., last-touch, first-touch, linear, time-decay, and U-shaped method) underestimate the contribution of the search channel and overestimate those of other channels, particularly the display channel. They are unable to take the baseline effect into consideration and thus are outperformed by other methods. Among all the methods, our graphical attribution methods have the smallest estimation errors, which demonstrates the advantage of our proposed graphical attribution methods in measuring channels' contribution to conversions.

\section{Real Application}
In this section, we apply the proposed methods to a real-world use case. The data are about paid conversions of an online subscription product of a Fortune 500 company within $4$ consecutive months. There are $2,887,657$ paths and $74,440$ conversions in total.

The touchpoints belong to $Z=4$ channels with more specific details. For the search channel, a branded search is a specific company or product being advertised while a non-branded search is a generic search result, not for a specific company or product.
For the social channel, an owned social click is within the company’s control and not paid for (e.g. a corporate LinkedIn post) and an earned social click is outside the company's control but not paid for either (e.g. a third party sharing a corporate LinkedIn post).
For the email channel, awareness means the ad is just trying to make a customer aware of products. A promotion email means that there is a discount-priced product being offered.
Call to action means the customer is already familiar with the product, and the ad contains a specific call to action (e.g. buy now).

\begin{figure}[H]
\centering
\includegraphics[width=11cm]{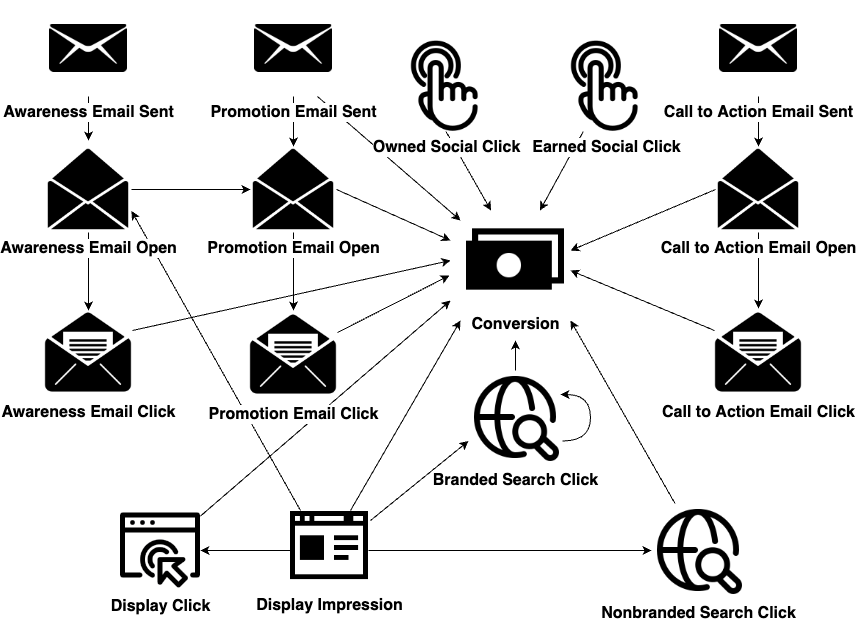}
\caption{The Granger causality graph for real data.}\label{realgraph}
\end{figure}

\begin{figure}[H]
\centering
\includegraphics[width=12cm]{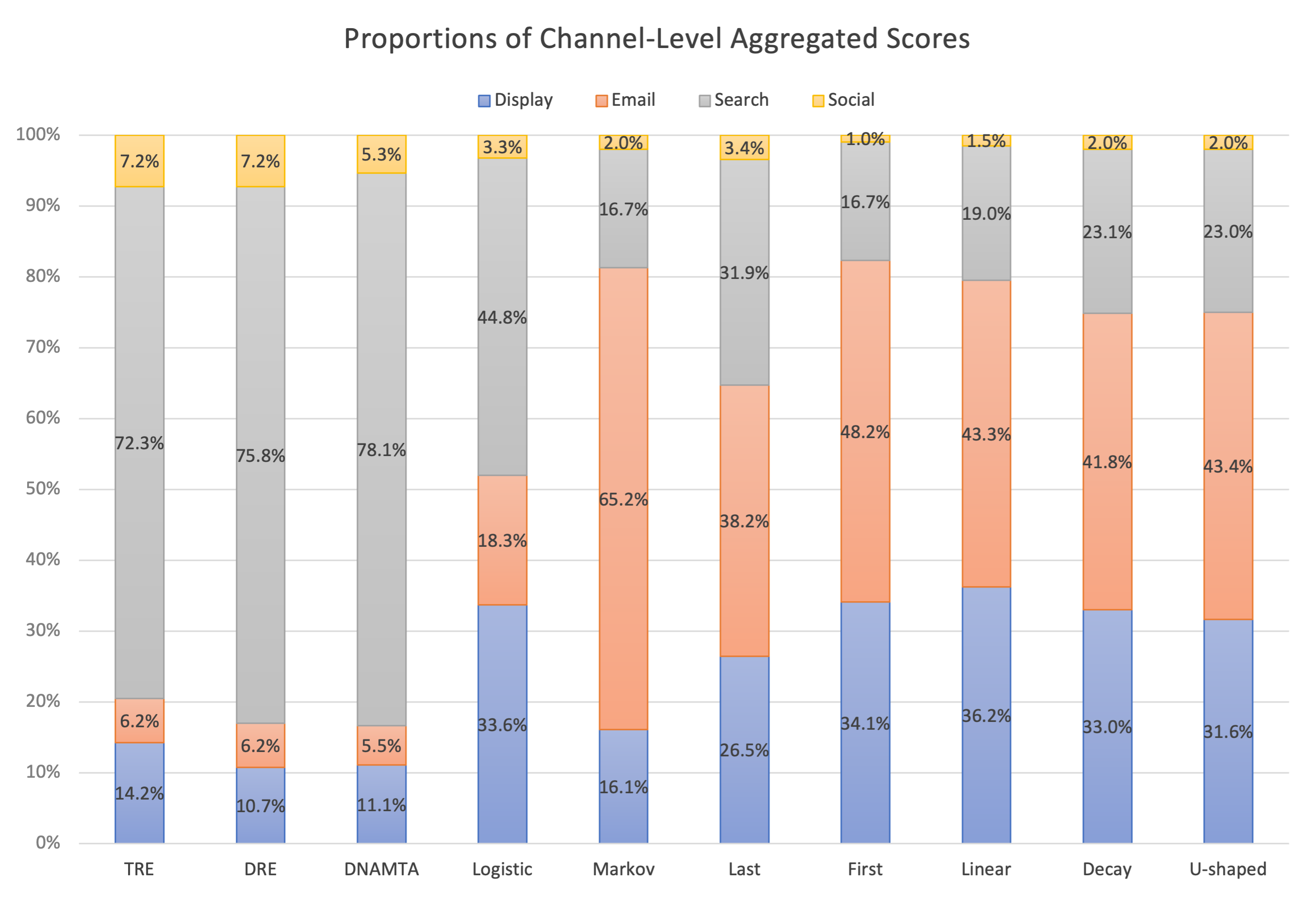}
\caption{The comparison of the proportions of CAS between graphical attribution methods and other methods. Each bar represents the proportions of CAS for an attribution method, with four channels colored differently in four parts. }\label{bar_real}
\end{figure}

The learned Granger causality graph is shown in Figure~\ref{realgraph}. Besides the excitation from touchpoints to a potential conversion, the graphical point process model finds the interactions between touchpoints within and across channels. For example, an awareness email open touchpoint may increase the chance of opening a promotion email. A display impression can trigger search clicks and awareness emails open. There is also a self-loop for the branded search click, meaning that clicks of this type tend to appear in clusters. Figure~\ref{bar_real} visualizes the proportions of CAS among different methods. The two graphical attribution methods and DNAMTA have similar results by giving the search channel the highest proportion of credit ($\ge 70\%$). As far as the display channel is concerned, compared with the direct removal effect ($10.7\%$), the total removal effect assigns more credit ($14.2\%$) since a display impression may trigger a search click. Logistic regression emphasizes the importance of the display channel more than other algorithmic methods by assigning scores to the display channel ($33.6\%$) close to the search channel ($44.8\%$). The rule-based methods, together with the Markov model, tend to give the highest scores to the email channel ($\ge 38\%$).

Based on the Granger causality graph, there indeed exists a hierarchical structure for the event types, and thus it is necessary to build a model with the response being more than just conversion. The channel-level aggregated scores obtained from our graphical methods reflect that the search channel is the most effective channel. The total removal effect emphasizes the importance of the display channel, which may play a key role in the early stage of a positive path.

\section{Conclusion}
In this paper, we propose a novel graphical point process framework for multi-touch attribution. First, we develop a graphical point process model to analyze customer-level path-to-purchase data. The graphical model utilizes the Granger causality to reveal the exciting effects among touches as well as the direct conversion effects of numerous types of touchpoints. Then, in the framework of the point process, we further propose graphical attribution methods to allocate proper conversion credit to individual touchpoints and the corresponding channels for each customer’s path to purchase. Our proposed attribution methods consider the attribution score as the removal effect, and we study two types of removal effects. We provide the probabilistic definition and the mathematical form of the removal effects. We develop a new efficient thinning-based simulation method and a backpropagation algorithm for the calculation. We employ a regularization method to select edges and estimate parameters simultaneously. We develop an ADMM to solve this optimization problem with desired computational efficiency and scalability. In addition, we
provide a theoretical guarantee by establishing the asymptotic theory for parameter estimates.

{
\bibliographystyle{agsm}
\bibliography{ref.bib}
}

\end{document}